\documentclass[10pt, aps,prx,floatfix,showpacs,twocolumn,longbibliography,superscriptaddress]{revtex4-2}
\usepackage[english]{babel}
\usepackage[normalem]{ulem}
\usepackage[utf8]{inputenc}
\usepackage[T1]{fontenc}
\usepackage{dcolumn}
\usepackage{bm}
\usepackage{color}
\usepackage{amssymb}
\usepackage{amsmath}
\usepackage{graphicx}
\usepackage{amsfonts}
\usepackage{slashed}
\usepackage{pstricks}
\usepackage{float}
\usepackage{multirow}
\usepackage{hyperref}
\usepackage{array}
\allowdisplaybreaks
\usepackage{dcolumn}
\usepackage{epsf}
\usepackage{float}
\usepackage{subcaption}
\usepackage[nice]{nicefrac}
\usepackage{tikz}
\usetikzlibrary{positioning}
\usepackage{cancel}
\usepackage{mathrsfs}
\begin{document}

\title{Hypernuclei with Neural Network Quantum States}

\author{Andrea Di Donna}
\affiliation{Dipartimento di Fisica, University of Trento, Via Sommarive 14, I-38123 Povo, Trento, Italy}
\affiliation{INFN-TIFPA Trento Institute of Fundamental Physics and Applications, Via Sommarive 14, I-38123 Povo, Trento, Italy}

\author{Lorenzo Contessi}
\affiliation{Universite Paris-Saclay, CNRS-IN2P3, IJCLab, 91405 Orsay, France}

\author{Alessandro Lovato}
\affiliation{Physics Division, Argonne National Laboratory, Argonne, Illinois 60439, USA}
\affiliation{Computational Science Division, Argonne National Laboratory, Argonne, Illinois 60439, USA}
\affiliation{INFN-TIFPA Trento Institute of Fundamental Physics and Applications, Via Sommarive 14, I-38123 Povo, Trento, Italy}

\author{Francesco Pederiva}
\affiliation{Dipartimento di Fisica, University of Trento, Via Sommarive 14, I-38123 Povo, Trento, Italy}
\affiliation{INFN-TIFPA Trento Institute of Fundamental Physics and Applications, Via Sommarive 14, I-38123 Povo, Trento, Italy}

\date{\today}
\begin{abstract}
Leveraging complementary machine-learning-based approaches, we compute properties of $s$- and $p$-shell $\Lambda$ hypernuclei --- including binding energies, single-particle densities, and radii --- starting from the individual interactions among their constituents. These interactions are modeled using an improved leading-order pionless effective field theory expansion, with coefficients determined via a Gaussian Process framework anchored on virtually exact few-body techniques. We solve the many-body Schr\"odinger equation using a variational Monte Carlo method based on neural network quantum states, extending it for the first time to include $\Lambda$ particles alongside protons and neutrons. The predicted binding energies show remarkably good agreement with experimental results, given the simplicity of the input Hamiltonian. We also confirm the experimentally observed shrinkage of the proton radius in $^7_\Lambda$Li compared to its parent nucleus, $^6$Li. This work paves the way for an ab initio description of medium-mass and heavy hypernuclei, as well as for understanding the onset of strange degrees of freedom in the core of neutron stars.
\end{abstract} 

\maketitle

\section{Introduction}
Microscopic approaches to nuclear structure have witnessed tremendous progress over the last two decades. Two- and three-nucleon interactions are derived systematically within nuclear effective field theories~\cite{Epelbaum:2008ga,Machleidt:2011zz,Hammer:2019poc}, and quantitative estimates of their uncertainties can be obtained by leveraging Bayesian methods~\cite{Svensson:2021lzs,Bub:2024gyz}. Using these interactions as input, quantum many-body approaches based on single-particle basis expansions --- including coupled-cluster theory~\cite{Hagen:2013nca}, the in-medium similarity renormalization group~\cite{Hergert:2015awm}, and Gorkov-Green’s function methods~\cite{Dickhoff:2004xx,Soma:2012zd} ---are now routinely used to predict and describe properties of medium-heavy nuclei with controlled approximations. Quantum Monte Carlo methods have also advanced significantly, providing high-resolution descriptions of static and dynamic properties of nuclei with up to $ A \simeq 20$ nucleons, including electromagnetic transitions, beta-decay rates, and electroweak response functions, with high accuracy~\cite{Carlson:2014vla,Lynn:2019rdt}. 

The frontier in nuclear physics now lies with exotic systems. A primary example is neutron-rich nuclei, which are experimentally studied at facilities such as FAIR, FRIB, JLab, J-PARC, and RIBF. These experiments provide invaluable information on the limits of nuclear stability and the abundances of elements. Another example is hypernuclei, self-bound systems composed of at least one hyperon in addition to protons and neutrons. The study of hypernuclei has a long experimental history: the first emulsion experiment was reported at a meeting of the Polish Academy of Sciences in 1952~\cite{Danysz:1953}. Since then, accelerator-based experiments have focused extensively on measuring ground-state properties, excitation energies, transition strengths, and charge symmetry breaking effects~\cite{May:1983oof,Tanida:2000zs,Akikawa:2002tm,Hashimoto:2006aw,J-PARCE13:2015uwb,Gogami:2015tvu,Gal:2016boi}.

Hypernuclear physics also plays a fundamental role in understanding the phenomenology of dense nuclear matter found in the interior of neutron stars~\cite{Bombaci:2016xzl,Maslov:2015msa}. At densities greater than twice nuclear saturation, the conversion of nucleons into hyperons may become energetically favorable, and Pauli blocking would prevent hyperons from decaying by limiting the phase space available to nucleons. Hyperon formation would lead to a reduction in the Fermi pressure exerted by the baryons and a softening of the equation of state, which would not be able to support the most massive neutron stars observed~\cite{Demorest:2010bx,Antoniadis:2013pzd,NANOGrav:2019jur,Fonseca:2021wxt}. This apparent inconsistency is a longstanding problem known as the {\it hyperon puzzle}~\cite{Lonardoni:2014bwa,Haidenbauer:2016vfq}. Its solution requires elucidating hyper-nuclear interactions through a combination of microscopic analysis and thorough experimental investigation.

For a long time, microscopic calculations of hypernuclei were limited to systems containing up to four baryons, relying on phenomenological potentials~\cite{Nogga:2001ef,Hiyama:2001zt,Nemura:2002fu,Filikhin:2002wp,Garcilazo:2007pc}. More recently, few-body techniques combined with pionless effective field theory (EFT) interactions have been successfully employed to address the long-standing overbinding problem in light $\Lambda$ hypernuclei~\cite{Contessi:2018qnz}. Quantum Monte Carlo methods have enabled a consistent description of a wide range of hypernuclei and neutron-star matter, though they still rely on somewhat simplified phenomenological interactions~\cite{Lonardoni:2013rm,Lonardoni:2013gta,Lonardoni:2014bwa}. Additionally, the no-core shell model has been applied to perform microscopic calculations of $s$- and $p$-shell hypernuclei, starting from two- and three-baryon interactions consistently derived within chiral EFT~\cite{Wirth:2014apa,Wirth:2016iwn,Wirth:2017lso,Wirth:2019cpp}. These calculations contribute to constraining the hyperon-nucleon interaction, using experimental ground-state and spectroscopic data for selected $p$-shell hypernuclei~\cite{Knoll:2023mqk}. Even more recently, Ref.~\cite{Le:2024rkd} reported the first calculation of light $\Lambda$ hypernuclei with up to $A=7$ based on consistent two and three body hypernuclear potentials, both derived within chiral EFT.

In this work, we combine complementary machine-learning-based approaches to compute the properties of $s$- and $p$-shell single-$\Lambda$ hypernuclei, paving the way for calculations of medium-mass systems with quantified uncertainties. Nuclear and hypernuclear forces are modeled using an improved leading-order pionless EFT expansion, with low-energy constants determined through a Gaussian Process framework informed by highly accurate few-body calculations. Using this Hamiltonian as input, we compute the ground-state properties of selected $s$- and $p$-shell $\Lambda$ hypernuclei by extending the state-of-the-art variational Monte Carlo method based on neural network quantum states (VMC-NQS)~\cite{Adams:2020aax,Lovato:2022tjh,Kim:2023fwy,Fore:2024exa} to include $\Lambda$ particles alongside protons and neutrons. We present results for binding energies, $\Lambda$-separation energies, and nuclear and hypernuclear radii and densities, including an analysis of the experimentally observed shrinkage of the proton radius in $^7_\Lambda$Li. Our work lays the foundation for accurate calculations of medium-mass and heavy hypernuclei, as well as hypernuclear matter, with implications for resolving the long-standing hyperon puzzle.

\section{Hamiltonian}
\label{sec:hamiltonian}
The starting point of our analysis is a non-relativistic Hamiltonian describing single-$\Lambda$ hypernuclei $^{A}_{\Lambda}Z$, where $A$ is the total number of baryons, which reads
\begin{equation}
H = -\sum_i \frac{\nabla_i^2}{2m_N} + V_{\text{NN}} + V_{\text{NNN}} - \frac{\nabla_\Lambda^2}{2m_\Lambda} + V_{\text{N}\Lambda} + V_{\text{NN}\Lambda}\,,
\end{equation}
where, Latin indices denote nucleons only, and we set $\hbar = 1$. The first three terms represent the Hamiltonian of the core nucleus $^{A}Z$, while the last three terms model the kinetic energy of the $\Lambda$ particle, the nucleon-$\Lambda$ (N$\Lambda$) potential, and the nucleon-nucleon-$\Lambda$ (NN$\Lambda$) potential.

The nucleon-nucleon (NN) potential is expressed as the sum of a charge-independent term and the electromagnetic interaction: \( V_{\rm NN} = V_{\rm NN}^{\rm CI} + V_{\rm NN}^{\rm EM} \). Following Ref.~\cite{Schiavilla:2021dun}, we assume \( V_{\rm NN}^{\rm EM} \) to include only the Coulomb repulsion between finite-size protons. We use the parametrization described in Ref.~\cite{auerbachTheoryIsobaricAnalog1972}, which accounts for finite-size effects through dipole form factors for the interacting protons. Other terms proportional to the first and second powers of the fine-structure constant—such as two-photon interactions, Darwin-Foldy corrections, vacuum polarization, and magnetic interactions—are neglected.

We model the strong components of both nuclear and hypernuclear forces in a consistent manner inspired by a leading-order pionless EFT expansion~\cite{Schiavilla:2021dun,Contessi:2018qnz}. The charge-independent components of the nucleon-nucleon and hyperon-nucleon potentials are given by 
\begin{align}
V_{NN}^{\rm CI} &= \sum_{\rm IS} C^{\text{IS}}_{\lambda} \sum_{i < j} \mathcal{P}_{\text{IS}}(ij)\delta_{\lambda}(\mathbf{r}_{ij}), \\
V_{\text{N}\Lambda} &= \sum_{\text{IS}} \tilde{C}^{\text{IS}}_{\lambda} \sum_{i} \mathcal{P}_{\text{IS}}\left(i\Lambda\right) \delta_{\lambda}(\mathbf{r}_{i\Lambda}),
  \label{eqn:LO_2b_potential}
\end{align}
where $\mathcal{P}_{\text{IS}}$ are projection operators for NN and N$\Lambda$ pairs with total isospin $I$ and spin $S$, respectively, while $C^{\text{IS}}_{\lambda}$ and $\tilde{C}^{\text{IS}}_{\lambda}$ are the corresponding two-body low-energy constants (LECs), whose values depend upon the regulator. We assume the presence of a single shallow virtual state in the $N-\Lambda$ system, restricted to each s-wave channel only, an approach that has proven successful in the past. The relevant isospin-spin combinations in the NN sector are \(\text{IS} = \{(10), (01)\}\), corresponding to the \(^1S_0\) and \(^3S_1\) channels. Since strong and electromagnetic interactions do not mix nucleons with \(\Lambda\) particles, the latter are distinguishable and can be taken to have isospin zero. Hence, the s-wave channels for the \(N \Lambda\) potential are $\text{IS} = \left\{\left(\frac{1}{2} 0\right), \left(\frac{1}{2} 1\right)\right\}$.

In coordinate space, contact interactions are typically regularized by introducing a local Gaussian regulator, specified by its cutoff $\lambda$:
\begin{equation}
  \delta_{\lambda}(\mathbf{r}) = \exp\left(-\frac{\lambda^2}{4} \mathbf{r}^2\right).
  \label{eqn:Gaussian_regulator}
\end{equation}

At leading order, the three-body interaction consists of a single spin-isospin independent NNN term and three NN$\Lambda$ contributions associated with the $\text{IS} = \left\{\left(0 \tfrac{1}{2}\right), \left(1 \tfrac{1}{2}\right), \left(0 \tfrac{3}{2}\right)\right\}$ channels, with explicit forms given by: 
\begin{align}
        \label{eqn:Vnnn}
         V_{\text{NNN}} &= D_{\lambda} \sum_{i < j < k} \left(\sum_{\text{cyc.}} \delta_{\lambda}(\vec{r}_{ik}) \delta_{\lambda}(\vec{r}_{jk}) \right), \\
        \label{eqn:Vlnn}
        V_{\text{NN}\Lambda} &= \sum_{\text{IS}} \tilde{D}^{\text{IS}}_{\lambda} \sum_{i < j} \mathcal{Q}_{\text{IS}}(ij\Lambda) \delta_{\lambda}(\vec{r}_{i\Lambda}) \delta_{\lambda}(\vec{r}_{j\Lambda}).
\end{align}
The sum in Eq.~\eqref{eqn:Vnnn} runs over all NNN triplets, where $\sum_{\text{cyc.}}$ denotes cyclic permutations among the particles within each triplet. Conversely, the sum in Eq.~\eqref{eqn:Vlnn} runs over all NN pairs, with the operators $\mathcal{Q}_{\text{IS}}$ projecting the interaction onto baryon triplets with specific isospin \(I\) and spin \(S\). Lastly, \(D^\lambda\) and \(\tilde{D}_{\lambda}^{\text{IS}}\) represent the LECs for NNN and NN$\Lambda$ interactions, respectively, for a given value of the regulator \(\lambda\) and spin-isospin channel. Note that the same regulator value is used across all three spin-isospin channels defining the NN$\Lambda$ potential. 

\begin{table*}[!htb]
\renewcommand{\arraystretch}{1.2} 
\caption{\label{tab:two_body_fit_parameters} Numerical values for the two-body LECs and regulators adjusted to reproduce the scattering lengths and effective ranges extracted from scattering experiments in the spin-singlet and spin-triplet channels. The np values are taken from Ref.~\cite{Hackenburg:2006qd}, and the p$\Lambda$ ones from Ref.~\cite{Alexander:1968acu}.}
\centering
\begin{tabular}{c|ccc|cc}
\hline
\textrm{Channel} & \textrm{IS} & $C_{\text{IS}}$ (MeV) & $\lambda_{\text{IS}}$ (fm$^{-1}$) & $a_{\text{exp}}$ (fm) & $r_{\text{exp}}$ (fm) \\
\hline
\multirow{2}{*}{np}  & $10$ & -31.0633 & 1.10117 & -23.7148(43) & 2.750(18) \\
                     & $01$ & -68.3747 & 1.30512 & 5.4112(15) & 1.7436(19)  \\
\hline
\multirow{2}{*}{p$\Lambda$}  & $\frac{1}{2}0$ & -33.5417 & 1.54720 & -1.80 & 2.80  \\
                     & $\frac{1}{2}1$ & -25.3115 & 1.41379 & -1.60 & 3.30 \\
\hline
\end{tabular}
\end{table*}

\subsection{Fitting procedure}
In pionless EFT, the independence of observables with respect to the regulator cutoff $\lambda$ is ensured by allowing the low-energy constants (LECs) to run with $\lambda$, and by taking the cutoff to be much larger than the breaking scale of the theory~\cite{Bedaque:2002mn,Hammer:2019poc}. However, standard applications of pionless EFT to many-body systems are severely limited by the instability of multi-fermion states at leading order~\cite{Stetcu:2006ey}. In particular, taking a too large cutoff renders $^{16}$O unstable with respect to breakup into four $^4$He nuclei~\cite{Contessi:2017rww,Bansal:2017pwn}. To overcome this issue, we adopt the strategy introduced in Refs.~\cite{Lu:2018bat,Schiavilla:2021dun,Gnech:2023prs}, which captures the essential elements of nuclear binding while preserving formal consistency. Instead of taking $\lambda$ to infinity, we fix its value separately in the NN and N$\Lambda$ channels to reproduce the experimentally extracted effective ranges. The regulators for the NNN and NN$\Lambda$ three-body forces are likewise tuned to experimental input, as discussed below. This approach is analogous to improving the convergence of pionless EFT by resumming perturbative subleading corrections, as recently proposed  in Refs.~\cite{Contessi:2023yoz,Contessi:2024vae,Contessi:2025xue} It can be systematically improved by including higher-order terms, while simultaneously restoring the renormalization group invariance that is otherwise lost when fixing the interaction range.

As in~\cite{Schiavilla:2021dun}, we determine the LECs and regulator values of the NN potential by fitting to the neutron–proton (np) scattering lengths and effective ranges in the ${}^1S_0$ and ${}^3S_1$ channels, corresponding to isospin states $\text{IS} = \{(10), (01)\}$, respectively. For the experimental input, we adopt the values reported in~\cite{Hackenburg:2006qd}. Similarly, the LECs and regulators for the N$\Lambda$ interaction are fixed to reproduce the scattering lengths and effective ranges inferred from low-energy $\Lambda p$ scattering cross sections measured in~\cite{Alexander:1968acu}, in both the ${}^1S_0$ and ${}^3S_1$ channels, which correspond to isospin states $\text{IS} = \{(\frac{1}{2}0), (\frac{1}{2}1)\}$. The numerical fits are carried out using the \textit{variable phase approach} ~\cite{calogeroNovelApproachElementary1963}. The results are summarized in Table~\ref{tab:two_body_fit_parameters}, which lists the fitted LECs, regulators, and corresponding scattering parameters. The LECs and regulators obtained for the $np$ channel are in close agreement with those reported in~\cite{Schiavilla:2021dun}, with only minor deviations arising from slight differences in the scattering lengths and effective ranges used in the fits.

\begin{table}[!b]
\caption{\label{tab:fitted_BE}%
Experimental ground-state energies used to constrain the three-body NNN interaction taken from~\cite{iaea_chart}. The uncertainties were inflated to $2.5$\% of their respective values.}
\centering
\renewcommand{\arraystretch}{1.2} 
\begin{tabular}{lcc}
\hline
\textrm{System} &
\multicolumn{1}{c}{$B_E$ (MeV)} &
\multicolumn{1}{c}{$\sigma$ (MeV)} \\
\hline
\(^3\text{H}\)   & 8.48  & 0.21  \\
\(^3\text{He}\)   & 7.80  & 0.20  \\
\(^4\text{He}\)  & 28.30 & 0.71  \\
\(^{16}\text{O}\) & 127.6 & 3.2  \\
\hline
\end{tabular}
\end{table}

In principle, the coupling constant \( D_\lambda \) and the regulator \( \lambda \) defining the NNN potential could be determined by fitting the ground-state energies of \( ^3\text{H} \) and \( ^4\text{He} \). 
However, owing to the strong correlation between these observables, many different combinations of LECs and regulators can reproduce them. This behavior is expected from renormalization group invariance and from the absence of a four-body force at leading order in the theory. Nevertheless, excessively large cutoffs would lead to instabilities in larger systems, we include the ground-state energy of \( ^{16}\text{O} \) as an additional fitting constraint. This choice remains consistent with the EFT framework, provided that the fixed range does not exceed the inverse of the theory’s breakdown scale. The experimental values and their uncertainties used in the fit are reported in Table~\ref{tab:fitted_BE}. To improve the stability of the fitting algorithm we have artificially inflated the uncertainties of the ground-state energies to \( 2.5\% \) of their respective values, well below the theoretical uncertainty associated with a leading-order pionless-EFT interaction~\cite{Ekstrom:2024dqr,Bub:2024gyz}. 

Similarly, the low-energy constants and regulators of the NN$\Lambda$ potential can be adjusted to reproduce the \( \Lambda \) separation energies, defined as the energy difference between the hypernucleus and its parent nucleus:
\[
B_{\Lambda} = E_{^{A-1}Z} - E_{^{A}_{\Lambda}Z},
\]
for \( ^3_\Lambda\text{H} \), \( ^4_\Lambda\text{H}_{S=0} \), \( ^4_\Lambda\text{H}_{S=1} \), and \( ^5_\Lambda\text{He} \). As in the nuclear case, we also include the separation energy of  $^{16}_\Lambda$O in the fit. The experimental values and their uncertainties are listed in Table~\ref{tab:fitted_BLambda}.

\begin{table}[!b]
\caption{\label{tab:fitted_BLambda}%
Experimental $\Lambda$ separation energies used to constrain the three-body NN$\Lambda$ interaction taken from the Hypernuclear Database~\cite{eckertChartHypernuclidesHypernuclear2023}.}
\centering
\renewcommand{\arraystretch}{1.4} 
\begin{tabular}{lccc}
\hline
\textrm{System} &
\multicolumn{1}{c}{$B_\Lambda$ (MeV)} &
\multicolumn{1}{c}{$\sigma$ (MeV)} &
\multicolumn{1}{c}{IS Dependence} \\
\hline
\(^3_{\Lambda}\text{H}\)   & 0.164  & 0.043 & $\left(0\frac{1}{2}\right)$ \\
\(^4_{\Lambda}\text{H}_{S=0}\) & 2.169  & 0.042 & $\left(0\frac{1}{2}\right)$, $\left(1\frac{1}{2}\right)$ \\
\(^4_{\Lambda}\text{H}_{S=1}\) & 1.081  & 0.046 & $\left(0\frac{1}{2}\right)$, $\left(1\frac{1}{2}\right)$, $\left(0\frac{3}{2}\right)$ \\
\(^5_{\Lambda}\text{He}\)  & 3.102  & 0.030 & $\left(0\frac{1}{2}\right)$, $\left(1\frac{1}{2}\right)$, $\left(0\frac{3}{2}\right)$ \\
\(^{16}_{\Lambda}\text{O}\)  & 13.00  & 0.089 & $\left(0\frac{1}{2}\right)$, $\left(1\frac{1}{2}\right)$, $\left(0\frac{3}{2}\right)$ \\
\hline
\end{tabular}
\end{table}

The fit for both the NNN and NN$\Lambda$ potentials is efficiently performed by leveraging Gaussian Processes (GP)~\cite{rasmussenGaussianProcessesMachine2008} to interpolate ground-state energies as continuous functions of the LECs and regulators. This approach is particularly advantageous when fitting the NN$\Lambda$ force, as the high dimensionality of the parameter space makes a direct grid search computationally prohibitive. First, we generate a dataset of binding energies and separation energies calculated at specific parameter values:
\begin{equation}
\Big\{\big(D_\lambda, \lambda\big)_{i}, B_E\big(D_\lambda, \lambda\big)_{i}\Big\} , 
\Big\{\big(\tilde{D}^{\rm IS}_{\tilde{\lambda}}, \tilde{\lambda}\big)_i, B_\Lambda\big(\tilde{D}^{\rm IS}_{\tilde \lambda}, \tilde{\lambda}\big)_i\Big\}
\end{equation}
where the two sets correspond to the NNN and NN$\Lambda$ potentials, respectively. We solve light nuclei and hypernuclei using the highly accurate Stochastic Variational Method (SVM)~\cite{suzukiStochasticVariationalApproach1998c}, as it enables fast and precise calculations for few-body systems. In contrast, the ground-state energies of $^{16}$O and $^{16}_\Lambda\text{O}$ are computed using the variational Monte Carlo method based on neural-quantum states (VMC-NQS)~\cite{Adams:2020aax,Lovato:2022tjh,Fore:2024exa}, described in detail in Section~\ref{sec:vmc_nqs}. Specifically, we employ the SVM to generate a set of 1000 training points for light nuclei and hypernuclei due to its low computational cost. For $^{16}$O and $^{16}_\Lambda\text{O}$, which are more computationally intensive, we generate 50 points each. These datasets are then used to train GP models, enabling smooth and analytic continuations of nuclear binding energies and hypernuclear separation energies $B_E\left(D_\lambda, \lambda\right)$ and $B_\Lambda(\tilde{D}^{\rm IS}_{\tilde{\lambda}}, \tilde{\lambda})$. 

The GP interpolation not only predicts energies at untested parameter values but also quantifies their uncertainties, enhancing the robustness of the fitting procedure.
The GP kernel imposes a prior on the mapping $(D,\lambda) \mapsto E$ by encoding our prior assumptions about the functional form of $E(D,\lambda)$. A GP model’s prior is fully determined by its second-order statistics via the kernel, which defines the pairwise covariance function between points in the input space. In our calculations, we use a combination of {\it rational quadratic} and {\it matern32} kernels. As noted in~\cite{Duvenaud2014}, combining two kernels is analogous to a logical OR operation: the resulting kernel will exhibit large values if either kernel independently has large values. The rational quadratic kernel models infinitely differentiable functions with varying degrees of smoothness across the input space, while the matern32 kernel with $\nu = 3/2$ models smooth functions that may include inflection points where the derivative changes abruptly.

To accommodate differing length scales along the various axes of the parameter space, we apply the automatic relevance determination technique~\cite{mackayBayesianNonLinearModeling1996a}. Our analysis suggests that the hypernuclear $\tilde{D}_{0\frac{1}{2}}$ LEC, unlike the other two, varies only within a narrow region of the parameter space close to zero. This observation stems from the fact that the binding energy of $^3_{\Lambda}\text{H}$ depends solely on $\tilde{D}^{0\frac{1}{2}}_\lambda$ in the three-body sector, and the related separation energy is small --- see Table \ref{tab:fitted_BLambda}. Notably, the average distance between the $\Lambda$ particle and the proton-neutron center of mass is about $10$ fm, as reported in \cite{alicecollaborationMeasurementLifetimeLambda2023a}, a typical characteristic of halo hypernuclei. The hyperparameters of the kernel are optimized by maximizing the log-marginal likelihood. The posterior GP is derived using Algorithm 2.1 from~\cite{rasmussenGaussianProcessesMachine2008}, as implemented in the Python package \texttt{GPy}~\cite{GPyGitHub}.

\begin{figure}[!htb]
    \centering
    \includegraphics[width=0.6\linewidth]{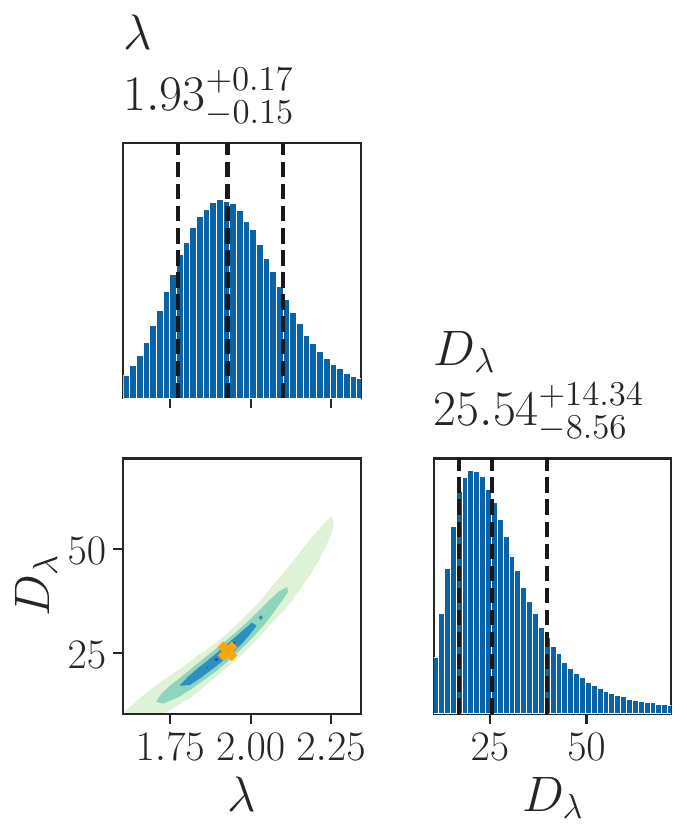}
    \caption{Joint distributions of the the NNN potential LEC (in MeV) and regulator (in fm$^{-1}$). The vertical dashed lines denote the 16th, 50th, and 84th percentiles each marginal distribution. The orange cross indicates the projection of the median point of the distribution.}
    \label{fig:correlation_lecs_nuclear}
\end{figure}

\begin{figure*}[!htb]
    \centering
    \includegraphics[width=0.6\linewidth]{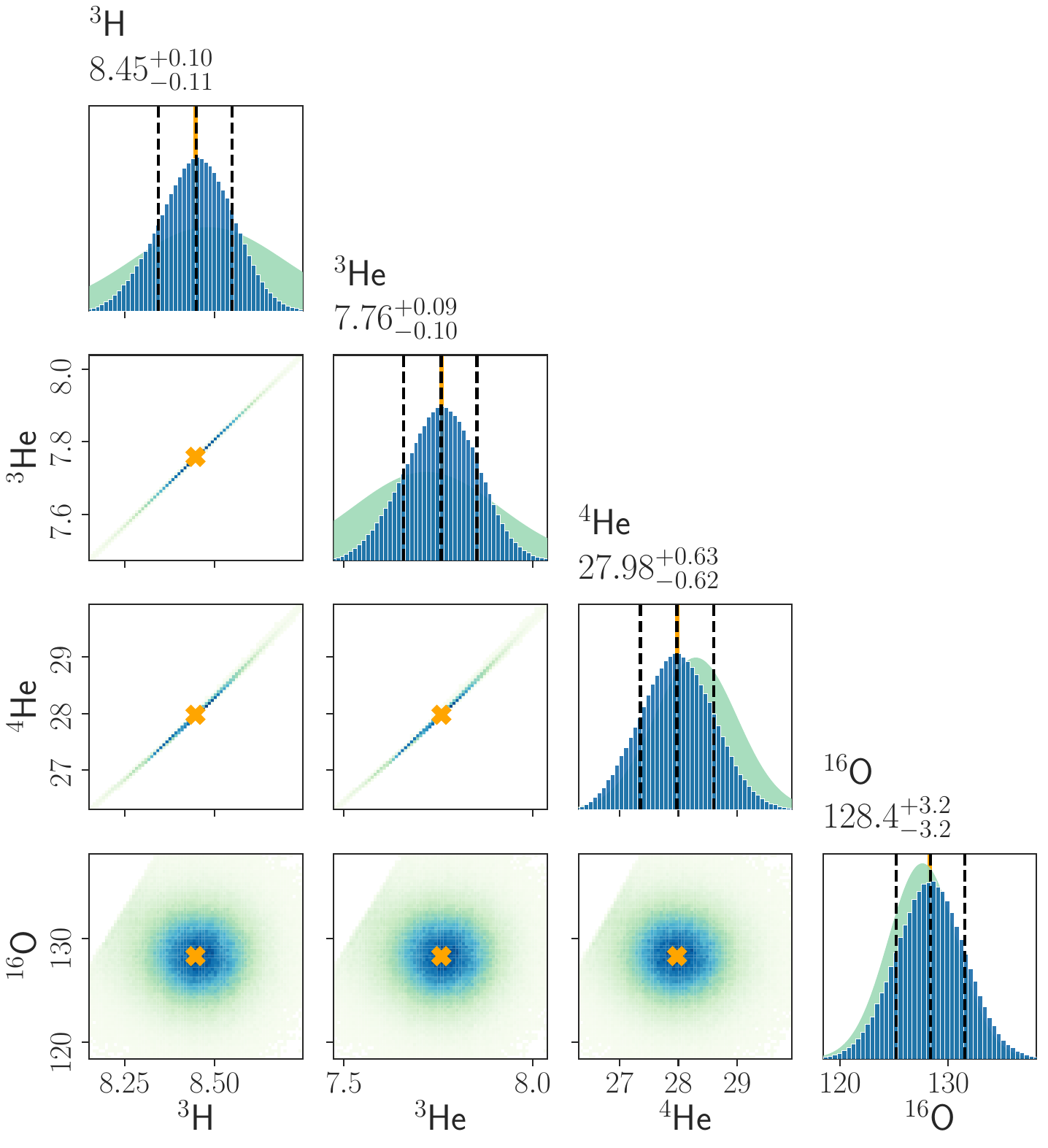}
    \caption{Joint posterior distributions of nuclear binding energies (MeV) obtained by varying the NNN potential LEC and regulator. Vertical dashed lines mark the 16th, 50th, and 84th percentiles of each marginal distribution. Orange markers and connecting lines indicate the binding energy computed with the median potential from Figure \ref{fig:correlation_lecs_nuclear}.}
    \label{fig:correlation_energies_nuclear}
\end{figure*}

\subsection{Sensitivity and Correlation Analysis}
Assuming uniform prior on the LECs and regulator values, the nuclear and hypernuclear LECs and regulator values are distributed according to a multivariate Gaussian likelihood
\begin{equation}
  \mathcal{N}(\vec{B} \vert \vec{D}, \lambda, \vec{\Sigma}) = \frac{e^{-\frac{1}{2} \left(\vec{B} - \vec{B}_{\mathrm{exp}}\right) \mathbf{\Sigma}^{-1} \left(\vec{B} - \vec{B}_{\mathrm{exp}}\right)}}{\sqrt{(2\pi)^k \det{\mathbf{\Sigma}}}} \, ,  
  \label{eqn:multivariate_gaussian_fit}
\end{equation}
where $\vec{B} = \vec{B}(\vec{D}, \lambda)$ is the vector of the binding energies or separation energies for the NNN and NN$\Lambda$ sector force, respectively, obtained through the GP. Analogously, $\vec{B}_{\mathrm{exp}}$ are the $k$ experimental values --- four in the NNN sector and five in the NN$\Lambda$ sector --- and $\mathbf{\Sigma}$ is the $k \times k$ diagonal covariance matrix corresponding to the experimental uncertainties. The nuclear and hypernuclear parameters distributions corresponds to $\mathcal{N}(\vec{B}_{E}\mid D_{\lambda},\lambda,\vec{\sigma})$ and $\mathcal{N}(\vec{B}_{\Lambda}\mid \tilde{D}_{\tilde{\lambda}},\tilde{\lambda},\vec{\sigma})$, respectively.

\begin{figure*}[!htb]
    \centering
    \includegraphics[width=0.6\linewidth]{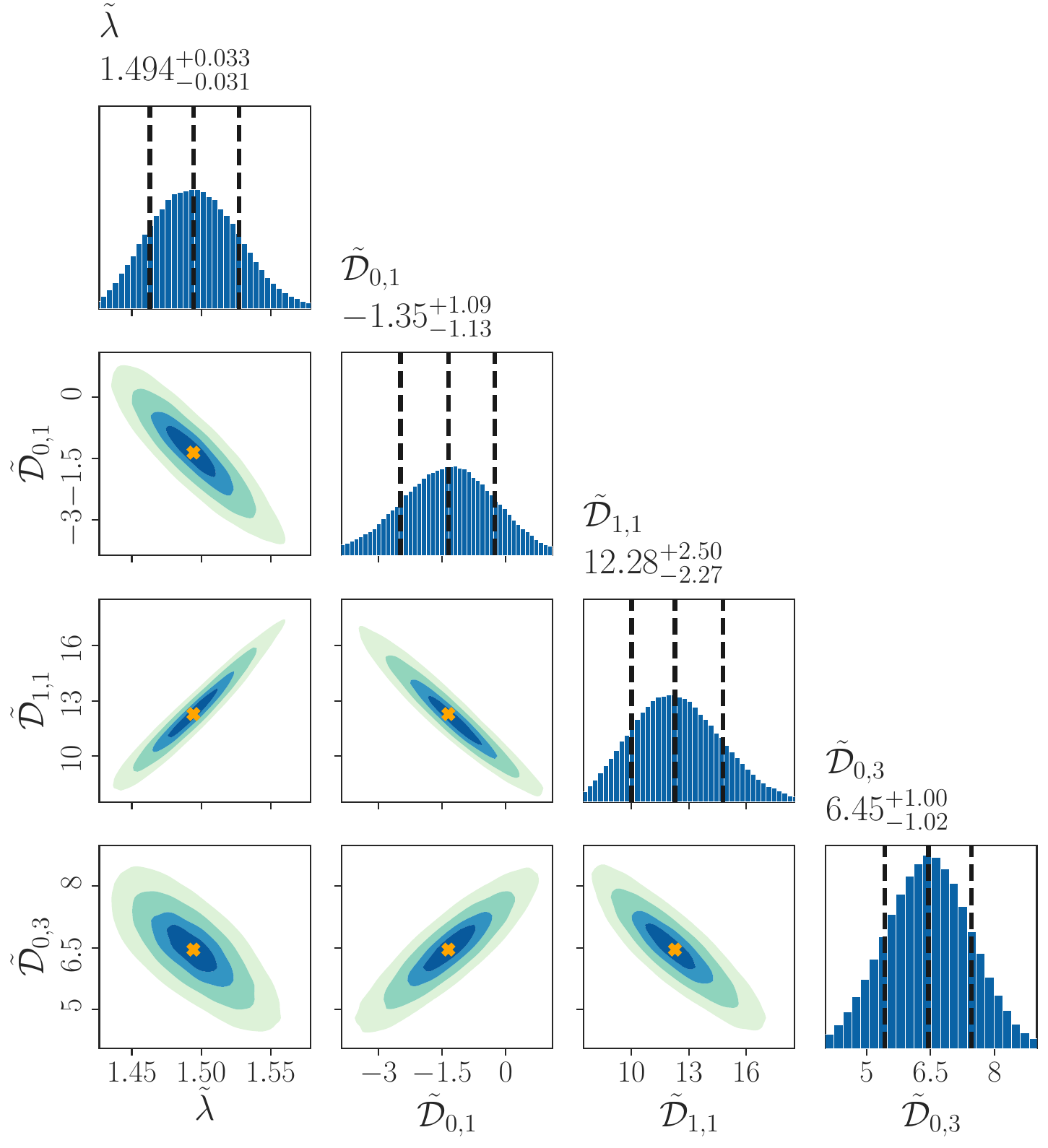}
    \caption{Joint distributions of the LECs (in MeV) and regulator (in fm$^{-1}$) entering NN$\Lambda$ potential.}
    \label{fig:correlation_lecs_hypernuclear}
\end{figure*}

\begin{figure*}[!htb]
    \centering
    \includegraphics[width=0.7\linewidth]{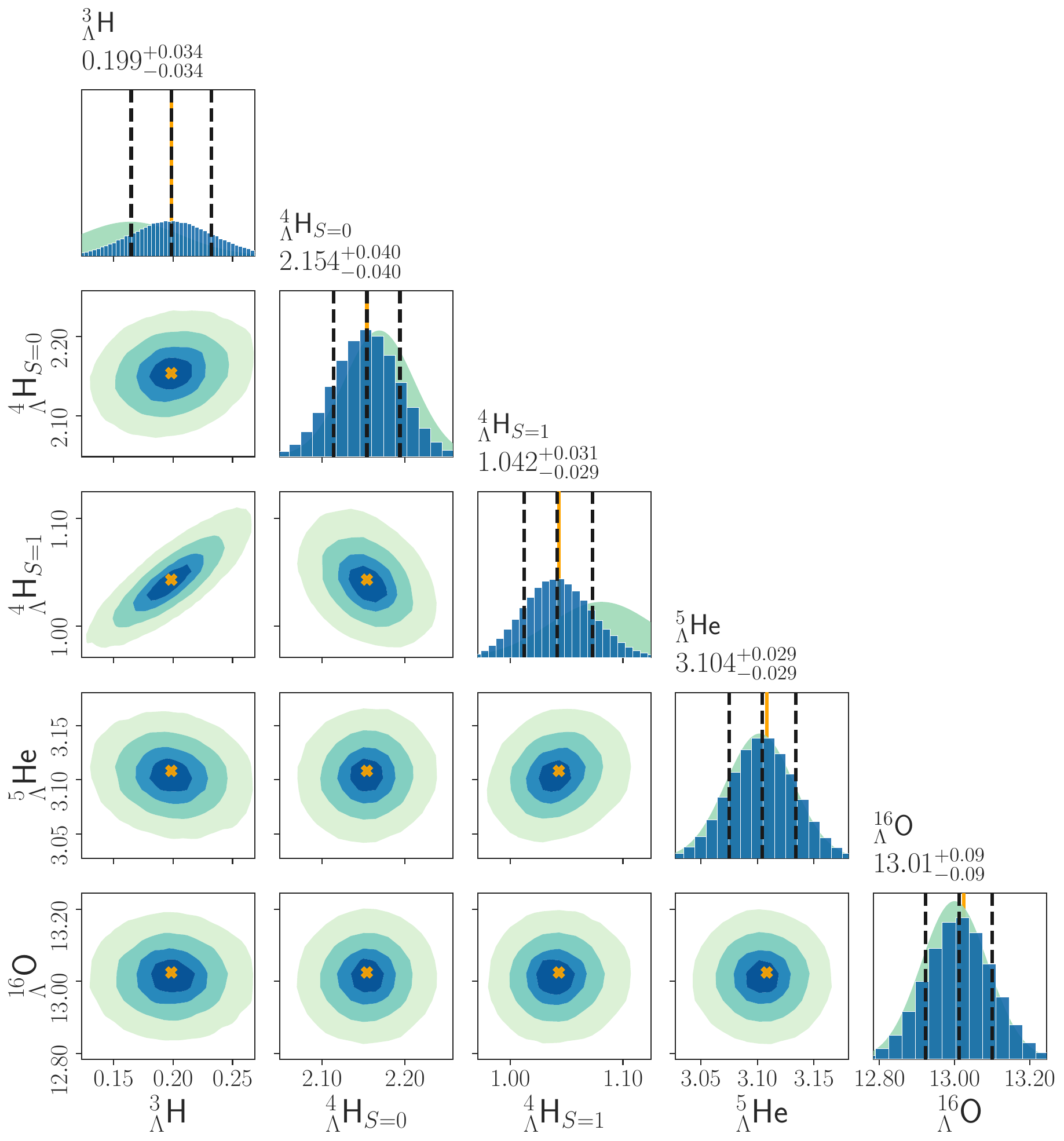}
    \caption{Same as Figure~\ref{fig:correlation_energies_nuclear} but for the hypernuclear separation energies (in MeV).}
    \label{fig:correlation_energies_hypernuclear}
\end{figure*}

To sample parameter distributions that accurately reflect experimental uncertainties, we employ Markov Chain Monte Carlo (MCMC) techniques—specifically, a robust sampler optimized for highly anisotropic distributions—as implemented in the \texttt{emcee} package~\cite{foreman-mackeyEmceeMCMCHammer2013}. This approach enables a thorough exploration of the parameter space, yielding statistically robust estimates of uncertainties and correlations among the LECs, the regulators, and predicted energies.
From the MCMC samples, we construct marginal and joint probability distributions for the LECs and regulators entering the NNN and NN$\Lambda$ potentials. In Figure~\ref{fig:correlation_lecs_nuclear}, we show corner plots displaying their marginal and joint posterior distributions for the NNN potential, while the joint posterior distributions of nuclear binding energies are shown Figure~\ref{fig:correlation_energies_nuclear}. The plots include histograms and credible intervals at the 16th, 50th, and 84th percentiles (corresponding to $\pm1\sigma$ for a normal distribution). The overlaying best-fit values and corresponding energies, are indicated by orange solid lines The distributions of the LECs and regulators entering the NN$\Lambda$ potential are displayed in Figure~\ref{fig:correlation_lecs_hypernuclear}, while the binding energies are illustrated in Figure~\ref{fig:correlation_energies_hypernuclear}. 

The joint posterior distributions highlight how correlations among observables emerge from the symmetries of the underlying Hamiltonian and propagate into relationships among the low-energy constants. In particular, the binding energies of \(^3\text{H}\), \(^3\text{He}\), and \(^4\text{He}\), shown in Figure~\ref{fig:correlation_energies_nuclear}, lie on an almost one-dimensional manifold in the \((\lambda, D_\lambda)\) plane, resulting in a strong positive correlation between these parameters. This behavior implies an approximate degeneracy when fitting only light nuclei, and the narrow ridge observed in Figure~\ref{fig:correlation_lecs_nuclear} extends to very large values of \(\lambda\). The inclusion of the heavier nucleus \(^ {16}\text{O}\) in the fit breaks this degeneracy and constrains \(\lambda\) to a finite interval. This effect is clearly visible in the same figure, where the strong correlations among the light nuclei are significantly mitigated by the addition of \(^ {16}\text{O}\). We note that the inclusion of \(^ {16}\text{O}\) in the calibration of three-body forces was first introduced in the construction of the chiral EFT interaction NNLO\(_\text{sat}\), which yields accurate binding energies and charge radii for nuclei up to \(^{40}\text{Ca}\)~\cite{Ekstrom:2015rta}.

The picture appears to be even richer in the NN$\Lambda$ sector. Each $B_\Lambda$ couples to a specific linear combination of the three spin--isospin channels: $\left(0 \tfrac{1}{2}\right)$, $\left(1 \tfrac{1}{2}\right)$, and $\left(0 \tfrac{3}{2}\right)$. Consequently, observables that probe different mixtures of these channels --- such as \(B_\Lambda(^4_\Lambda\text{H})\) versus \(B_\Lambda(^5_\Lambda\text{He})\) --- populate well-defined bands in the multidimensional space of LECs and range. Experimental uncertainties thus map directly onto the width of the allowed parameters' region: the more precise the \(B_\Lambda\) data, the narrower the credible intervals for the corresponding couplings and ranges. The joint posterior for \(B_\Lambda(^3_\Lambda\text{H})\) and \(B_\Lambda(^4_\Lambda\text{H}_{S=1})\) is markedly narrower than the individual experimental error bars reported in Table~\ref{tab:fitted_BLambda}. It is important to note that this tightening does not necessarily reflect an improved theoretical constraint on the experimental measurements, but rather the limited flexibility of our leading-order interaction model, which has to be verified upon inclusion of subleading orders. Although not as pronounced as in the NNN case, correlations among few-body observables in the hypernuclear sector broaden the convergence region of the fitted parameters into a continuous path in parameter space, which allows for very large values of $\lambda$. As in the nuclear case, including $^{16}_{\Lambda}\text{O}$ in the fit is essential to constrain the range of the NN$\Lambda$ interaction. This observable breaks the remaining degeneracy and selects a relatively narrow region in parameter space consistent with the experimental $B_\Lambda$ data.

\section{Hypernuclear neural network quantum state ansatz}
\label{sec:vmc_nqs}
We solve the nuclear quantum many-body problem by extending the highly flexible Pfaffian-Jastrow ansatz, originally introduced to model ultra-cold Fermi gases in Ref.~\cite{Kim:2023fwy} and later generalized to nuclear systems in Ref.~\cite{Fore:2024exa}, to include hyperons in addition to protons and neutrons. To clarify the notation, we denote the full set of coordinates as
\begin{equation}
X = (\mathbf{x}_1, \dots, \mathbf{x}_{A-1}, \mathbf{x}_{\Lambda}),
\end{equation}
where each $\mathbf{x}_i = (\mathbf{r}_i, \mathbf{s}_i)$ comprises the spatial Cartesian coordinates $\mathbf{r}_i$ and the $z$-components of the spin–isospin degrees of freedom, $\mathbf{s}_i = (s_{z_i}, t_{z_i})$, for the \(i\)-th particle. As a convention, we take $t_{z} = 1$ for protons, $t_{z} = -1$ for neutrons, and $t_{z} = 0$ for the $\Lambda$ particle. 

We first focus on the pure nucleonic sector and collectively denote the coordinates of the nucleons by
\begin{equation*}
X_N = (\mathbf{x}_1, \dots, \mathbf{x}_{A-1}).
\end{equation*}
The amplitude of the Pfaffian-Jastrow ansatz for the nucleonic sector can be schematically written as
\begin{align}
\langle X_N | \Psi_{PJ}\rangle \equiv \Psi_{PJ}(X_N) = e^{J(X_N)} \times \operatorname{Pf}\big[\Phi(X_N)\big]\,,
\label{eqn:PJ_Nuc_ansatze}
\end{align}
where the complex-valued, permutation-invariant Jastrow factor is defined as
\begin{equation}
J(X_N) = a\,\tanh\left(\frac{U_J(X_N)}{a}\right) + i\,V_J(X_N)\,.
\label{eq:reg_jas}
\end{equation}
Here, \(U_J(X_N)\) and \(V_J(X_N)\) are real-valued functions representing the logarithmic amplitude and phase, respectively. The parameter \(a\) acts as a cutoff that regularizes the growth of \(U_J\) and helps mitigate potential runaway instabilities. Following Ref.~\cite{bukov:2021}, we choose \(a = 8\), which allows a maximum relative magnitude variation of approximately \(10^{-7}\). We enforce both $U_J(X_N)$ and $V_J(X_N)$ to be permutation invariant by employing the Deep Sets architecture~\cite{Zaheer:2017,Wagstaff:2019} with {\it logsumexp} pooling:
\begin{align}
    U_J(X_N)&= \rho_U\left[\log\left(\sum_{i\neq j = 1}^{A-1} \exp\big({\phi_U(\mathbf{x}_i, \mathbf{x}_j)}\big)\right)\right]\nonumber\\
    V_J(X_N)&= \rho_V\left[\log\left(\sum_{i\neq j = 1}^{A-1} \exp\big({\phi_V(\mathbf{x}_i, \mathbf{x}_j)}\big)\right)\right]\,,
\end{align}
where $i$ runs over all the NN pairs. Both $\phi_{U,V}$ and $\rho_{U,V}$ are dense feed-forward neural networks, each comprising two hidden layers with 32 nodes. The latent space---the output dimension of $\phi_{U,V}$ and the input dimension of $\rho_{U,V}$---is set to 16.

The antisymmetric part of the nucleonic wave function is given by the Pfaffian of a skew-symmetric matrix, which has proven effective in capturing strong pairing correlations in both ultra-cold Fermi gases~\cite{Kim:2023fwy} and low-density nuclear matter~\cite{Fore:2024exa}. For nuclei with an even number of nucleons, we define the skew-symmetric matrix as
\begin{equation}
   \Phi(X_N)= \begin{bmatrix}
0 & \phi(\mathbf{x}_1, \mathbf{x}_2) & \dots & \phi(\mathbf{x}_1, \mathbf{x}_{A-1}) \\
\phi(\mathbf{x}_2, \mathbf{x}_1) & 0 & \dots & \phi(\mathbf{x}_2, \mathbf{x}_{A-1}) \\
\vdots & \vdots & \ddots & \vdots \\
\phi(\mathbf{x}_{A-1}, \mathbf{x}_1) & \phi(\mathbf{x}_{A-1}, \mathbf{x}_2) & \dots & 0
\end{bmatrix}\,.
    \label{eq:wavefunction_pfaffian}
\end{equation}
To ensure skew-symmetry, the pairing orbital is written as
\begin{equation}
    \phi(\mathbf{x}_i, \mathbf{x}_j)= 
    \eta(\mathbf{x}_i, \mathbf{x}_j) - \eta(\mathbf{x}_j, \mathbf{x}_i),
    \label{eq:skew_symmetric_elements}
\end{equation}
so that $\phi(\mathbf{x}_i, \mathbf{x}_j) = -\phi(\mathbf{x}_j, \mathbf{x}_i)$. In the above equation, $\eta(\mathbf{x}_i, \mathbf{x}_j)$ is a complex-valued function, whose logarithm is regularized as in Eq.~\eqref{eq:reg_jas} 
\begin{equation}
    \eta(\mathbf{x}_i, \mathbf{x}_j)=\exp\left[a\,\tanh\left(\frac{u_\eta(\mathbf{x}_i, \mathbf{x}_j)}{a}\right)+i \pi\,v_\eta(\mathbf{x}_i, \mathbf{x}_j))\right]\, .
\end{equation}
The real-valued feed-forward neural networks $u_\eta$ and $v_\eta$, which encode the logarithmic amplitude and phase of $\eta$, each include two hidden layers with $16$ nodes, and their outputs are one-dimensional.

Nuclei with an odd number of nucleons are treated by adding to the Pfaffian an unpaired single-particle orbital~\cite{Kim:2023fwy}. Thus, we extend $\Phi(X_N)$ to a $(2N+1)\times (2N+1)$ skew-symmetric matrix $\tilde{\Phi}(X_N)$ by introducing an additional row and column:
\begin{equation}
    \tilde{\Phi}= \begin{bmatrix}
 \Phi(X_N) & \mathbf{u} \\
-\mathbf{u}^T & 0
\end{bmatrix}\,,
    \label{eq:wavefunction_pfaffian_odd}
\end{equation}
where the vector $\mathbf{u} \in \mathbb{C}^N$ is defined as
\begin{equation}
     \mathbf{u}=\begin{bmatrix}
 \psi(\mathbf{x}_1) \\
\psi(\mathbf{x}_2)\\
\vdots \\
\psi(\mathbf{x}_N)
\end{bmatrix}\,.
\end{equation}
Consistent with the complex-valued functions above, we use two separate real-valued neural networks to parameterize the logarithm of $\psi(\mathbf{x}_i)$:
\begin{align}
    \psi(\mathbf{x}_i)=\exp\left[a\,\tanh\left(\frac{u_\psi(\mathbf{x}_i, \mathbf{x}_j)}{a}\right)+i \pi\,v_\psi(\mathbf{x}_i, \mathbf{x}_j)\right]\,.
    \label{eq:psi_single}
\end{align}
We again set \(a = 8\) to mitigate potential runaway instabilities. Both $u_\psi$ and $v_\psi$ are feedforward architectures with two hidden layers containing 16 nodes each, and their outputs are one-dimensional.

Since strong and electromagnetic interactions do not mix nucleons with $\Lambda$ particles, the latter can be assumed to be distinguishable, removing the requirement for the total wave function to be antisymmetric under exchange $\mathbf{x}_i \leftrightarrow \mathbf{x}_\Lambda$. Hence, in analogy with the separation of spin-up and spin-down electrons in the presence of purely Coulomb interactions the total wave function of single-$\Lambda$ hypernuclei can be written as
\begin{equation}
\langle X | \Psi_{PJ}^{\Lambda}\rangle \equiv \Psi_{PJ}^{\Lambda}(X) =  e^{J(X)} \times \psi_{\Lambda}(\mathbf{x}_{\Lambda}) \times \operatorname{Pf}\big[\Phi(X_N)\big]\,,
\label{eqn:Tot_PJH_ansatze}
\end{equation}
where $\psi_{\Lambda}(\mathbf{x}_{\Lambda})$ is a generalized single-particle orbital for the $\Lambda$ particle, whose logarithm is parameterized as in Eq.~\eqref{eq:psi_single}. To incorporate correlations among the nucleons and the $\Lambda$, the Jastrow factor now depends on the coordinates of all particles, and the summation in Eq.~\eqref{eq:reg_jas} includes both NN and N$\Lambda$ pairs. Assigning the $\Lambda$ particle an isospin projection value of $t_{z_\Lambda} = 0$, distinct from those of protons and neutrons, eliminates the need for a separate correlation function for nucleon–$\Lambda$ pairs within the Jastrow factor.

\subsection{Backflow Transformations}
To enhance the expressiveness of the ansatz, we apply a backflow transformation to the single-particle coordinates. This idea dates back to Ref.~\cite{Feynman:1956zz} and has recently emerged as a powerful and widely adopted strategy to augment the expressivity of anti-symmetric NQS~\cite{Luo:2019iaq,Hermann:2020xqs,Pfau:2020}. Following Ref.~\cite{Fore:2024exa}, we employ the message-passing neural network developed in Refs.~\cite{Pescia:2023mcc,Kim:2023fwy}, which has proven to efficiently capture correlations in both the homogeneous electron gas and ultra-cold Fermi gases.

We take the input single-particle ``visible'' features as \( \mathbf{v}_i = (\bar{\mathbf{r}}_i, \mathbf{s}_i) \). Note that—unlike in periodic systems, where only the spin and isospin of particle \( i \) are used as inputs—here the Cartesian coordinates of the nucleons are also included. To automatically remove spurious center-of-mass contributions from all observables~\cite{Massella:2018xdj}, we define the intrinsic spatial coordinates as \( \bar{\mathbf{r}}_i = \mathbf{r}_i - \mathbf{R}_{\rm CM} \), where \( \mathbf{R}_{\rm CM} \) denotes the center of mass of the nucleus (or hypernucleus). The ``visible'' pairwise features encode spatial as well as spin-isospin coordinates and are defined by \( \mathbf{v}_{ij} = [(\mathbf{r}_i-\mathbf{r}_j),\sqrt{(\mathbf{r}_i-\mathbf{r}_j)^2},\mathbf{s}_i,\mathbf{s}_j]\).

The initial hidden features for the nodes and the edges are obtained by concatenating the original and transformed single-particle and two-particle features, respectively, as
\begin{equation*}
\mathbf{h}_i^0 = [\mathbf{v}_i,  f_A(\mathbf{v}_i)]\quad, \quad \mathbf{h}_{ij}^0 = [\mathbf{v}_{ij}, f_B(\mathbf{v}_{ij})]\,.
\end{equation*}
The main role of the two fully connected neural networks, \( f_A \) and \( f_B \), is to preprocess the input coordinates. They also facilitate the implementation of the MPNN by ensuring that the dimensionality of the hidden features \( \mathbf{h}_i^t \) and \( \mathbf{h}_{ij}^t \) remains independent of \( t \). These two networks each consist of two fully connected hidden layers with $22$ nodes each. 

The MPNN update is performed iteratively for \( t = 0, \dots, T{-}1 \) layers. Each time, information between the node and edge features is exchanged through a so-called message
\begin{equation*}
\mathbf{m}_{ij}^t = f_M^t( \mathbf{h}_i^{t-1},  \mathbf{h}_{ij}^{t-1},  \mathbf{h}_j^{t-1})
\end{equation*}
For each particle \( i \), the relevant messages are collected and pooled to eliminate any ordering with respect to the other particles \( j \neq i \). As in Ref.~\cite{Kim:2023fwy}, we use logsumexp pooling, a smooth alternative to max pooling:
\begin{equation*}
\mathbf{m}_{i}^t =  \log\left(\sum_{j\neq i} \exp\left(\mathbf{m}_{ij}^t\right) \right)\,.
\end{equation*}
The hidden node and edge features are then updated as
\begin{align*}
\mathbf{h}_i^t &= \left(\mathbf{v}_i, f_F^t\left(\mathbf{h}_i^{t-1}, \mathbf{m}_{i}^t\right)\right)\,, \\
\mathbf{h}_{ij}^t &= \left(\mathbf{v}_{ij}, f_G^t\left(\mathbf{h}_{ij}^{t-1}, \mathbf{m}_{ij}^t\right)\right)\,.
\end{align*}
The functions \( f_M^t \), \( f_F^t \), and \( f_G^t \) are distinct fully connected neural networks with output dimensions matching those of \( f_A \) and \( f_B \), and, like the latter, consist of two fully connected hidden layers with 22 nodes each. Including concatenated skip connections to the visible features ensures that the signal from the raw input remains accessible even as the MPNN depth \( T \) increases. 

Finally, after the \( T \)-th iteration, we aggregate the hidden node and edge features into single-particle and pairwise feature vectors:
\begin{align*}
\mathbf{g}_{ij} &= \left(\mathbf{h}_i^T, \mathbf{h}_j^T, \mathbf{h}_{ij}^T \right)\,, \\ 
\mathbf{g}_i &=  \log\left(\sum_{j \neq i} \exp\left(\mathbf{g}_{ij}\right) \right)\,.
\end{align*}
By construction, \( \mathbf{g}_i \) is permutation equivariant with respect to the original input \( \mathbf{x}_i \), meaning that it depends on \( \mathbf{x}_i \) and is invariant under exchanges \( \mathbf{x}_j \leftrightarrow \mathbf{x}_k \) for all \( j, k \ne i \). Similarly, \( \mathbf{g}_{ij} \) is invariant under exchanges \( \mathbf{x}_l \leftrightarrow \mathbf{x}_m \) for all \( l, m \ne i, j \).

The pairwise feature vector \( \mathbf{g}_{ij} \) serves as input to both the Pfaffian and the Jastrow by replacing the original pair coordinates \((\mathbf{x}_i, \mathbf{x}_j)\) with \( \mathbf{g}_{ij} \). Note that the Pfaffian includes only nucleon–nucleon pairs, while the Jastrow also includes nucleon–$\Lambda$ pairs. Additionally, the permutation-equivariant output feature \( \mathbf{g}_\Lambda \) is used as input to the $\Lambda$ single-particle orbital $\Psi_\Lambda$.

\subsection{Benchmark with SVM}
The purely nucleonic Pfaffian-Jastrow ansatz of Eq.~\eqref{eqn:PJ_Nuc_ansatze} was recently validated by benchmarking the ground-state energies of selected nuclei with up to \( A = 6 \) nucleons, computed using the SVM method~\cite{Contessi:2025xue}. Agreement at better than the percent level was found using as input an improved-action pionless EFT Hamiltonians across a wide range of regulator values.

\begin{table}[!b]
\centering
\renewcommand{\arraystretch}{1.2}
\caption{Hypernuclear binding energies (in MeV) computed using the VMC algorithm with the Pfaffian-Jastrow NQS, compared with results from the SVM method. Both methods use as input the median NNN and NN$\Lambda$ potentials from Figures~\ref{fig:correlation_lecs_nuclear} and~\ref{fig:correlation_lecs_hypernuclear}.}
\label{tab:BLambda_comparison}
\begin{tabular}{lcc}
\hline
System & NQS & SVM \\
\hline
\(^3_{\Lambda}\text{H}\)            & 2.399(2) & 2.409 \\
\(^4_{\Lambda}\text{H}_{S=0}\)      & 10.606(3) & 10.601 \\
\(^4_{\Lambda}\text{H}_{S=1}\)      & 9.529(4) & 9.490 \\
\(^5_{\Lambda}\text{He}\)           & 31.086(3) & 31.082 \\
\hline
\end{tabular}
\end{table}

Here we focus on the hypernuclear sector. To gauge the accuracy of the NQS ansatz in Eq.~\eqref{eqn:Tot_PJH_ansatze}, we performed benchmark calculations using the SVM method for light hypernuclei, finding excellent agreement, as shown in Table~\ref{tab:BLambda_comparison}. In both methods, we use as input the median NNN and NN$\Lambda$ potentials from Figures~\ref{fig:correlation_lecs_nuclear} and~\ref{fig:correlation_lecs_hypernuclear}, respectively.

Note that to accelerate convergence to the ground-state energy, we enforce that the NQS be an eigenstate of the parity operator by symmetrizing the wave function as \( \Psi_{PJ}^{\Lambda}(X) \pm \Psi_{PJ}^\Lambda(\mathcal{P}X) \), where the $\mathcal{P}$ flips the spatial coordinates of all the particles, including the $\Lambda$
\begin{equation*}
\mathcal{P}(\mathbf{r}_1,\dots,\mathbf{r}_{A-1}, \mathbf{r}_\Lambda) = (-\mathbf{r}_1,\dots,-\mathbf{r}_{A-1}, -\mathbf{r}_\Lambda)
\end{equation*}

\begin{figure}[!htb]
\centering
\includegraphics[width=0.495\textwidth]{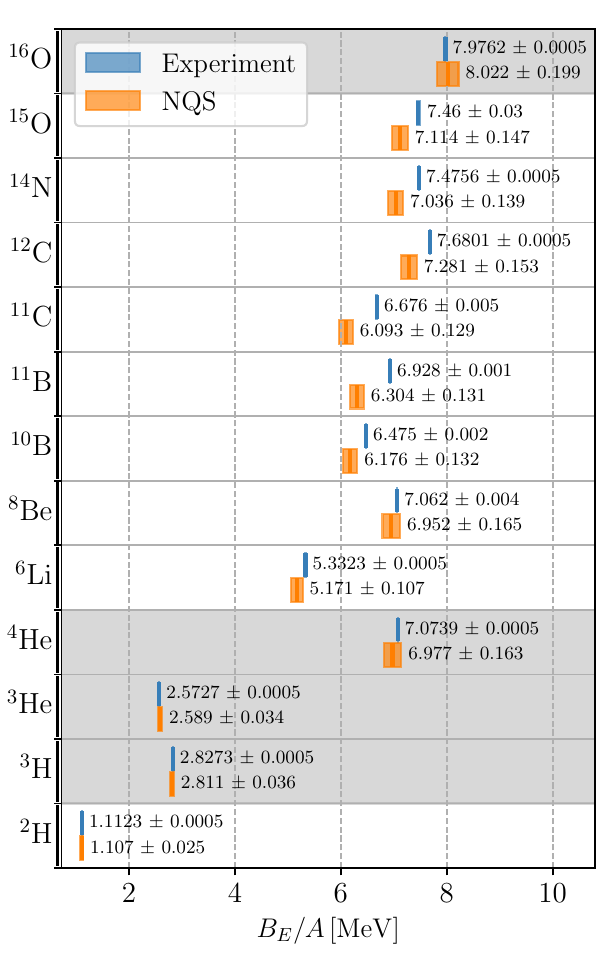}
\caption{Binding energy per nucleon for selected nuclei up to $^{16}$O. NQS results (orange) are compared with experimental data (blue). The shaded bands reflect both Monte Carlo statistical errors and model uncertainties in the coupling and range of the NNN force. Grey regions highlight the nuclei used to constrain the three-body force parameters. \label{fig:energies_nuclei}}    
\end{figure}

\section{Results}
To set the stage, Figure~\ref{fig:energies_nuclei} displays the binding energy per nucleon 
computed for selected nuclei that serve as a baseline for hypernuclear calculations. 
The shaded gray regions highlight the nuclei used to determine the $\lambda$ and $D_\lambda$ 
parameters of the NNN potential, as discussed in Section~\ref{sec:hamiltonian}. The uncertainties in the NQS calculations reflect both the statistical error from the variational VMC, $\sigma_{\text{VMC}}$, and the uncertainty associated with the NNN force, $\sigma_{\text{NNN}}$. The latter is estimated by sampling ten different realizations of the NNN potential, with $\lambda$ and $D_\lambda$ drawn from the joint posterior distribution shown in Figure~\ref{fig:correlation_lecs_nuclear}. As a conservative estimate, the total energy uncertainty is computed by summing the two contributions in modulus rather than in quadrature: $\sigma_{B_E} = \sigma_{\text{NQS}} + \sigma_{\text{NNN}}$. In principle, the truncation error of the theory should also be included. However, due to the improved nature of the interaction, a study of cutoff variations is not feasible, and the lack of experimental data further complicates an uncertainty estimate based on variations of the theory fit input.

Consistent with the findings of Refs.~\cite{Schiavilla:2021dun,Gnech:2023prs}, fitting the $np$ scattering length and effective range in the $^3S_1$ channel constrains the NN potential to accurately reproduce the deuteron binding energy. As for $^3$H, $^4$He, and $^{16}$O, the excellent agreement with experiment is expected, as their binding energies were used to determine the coupling and range of the NNN force. It is important to note, however, that the NQS binding energies, computed using a simple potential inspired by a leading-order pionless EFT expansion, successfully reproduce experimental values across several light nuclei—consistent with the findings of Ref.~\cite{Gnech:2023prs}. For the nuclei not included in the fit, we observe an average deviation of approximately 5\% between the calculated $B_E$ and experimental data. The largest discrepancies, around 9\%, occur for $^{11}$B and $^{11}$C, whose structure is strongly influenced by $p$-shell orbitals. This underbinding is likely due to the absence of explicit $p$-wave terms in the interaction~\cite{Gattobigio:2019omi}. Notably, the differences between theory and experiment are much smaller than the $\sim$30\% uncertainty typically expected at leading order in pionless EFT~\cite{Konig:2019xxk}. This improved agreement is likely due to the inclusion of finite-range effects in both the NN and NNN forces, which effectively resum higher-order contributions in the EFT expansion.

\begin{figure}[!htb]
    \centering
    \includegraphics[width=0.5\textwidth]{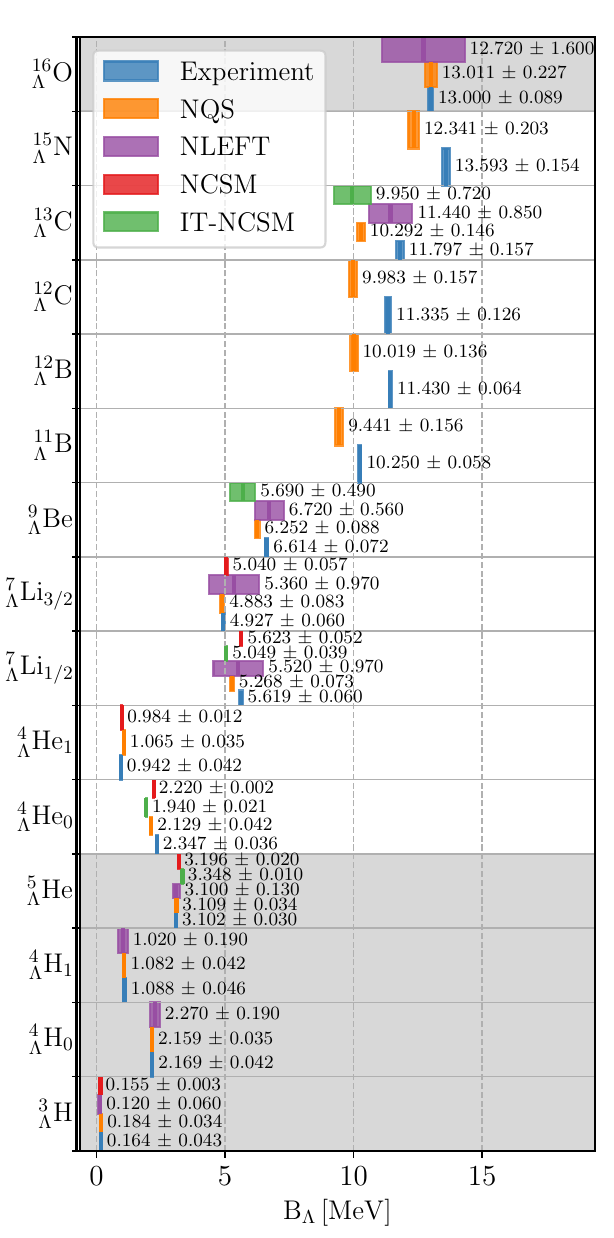}
    \caption{$\Lambda$-separation energies for selected hypernuclei up to $^{16}_\Lambda$O. NQS results (orange) are compared with experimental data from the Hypernuclear Database (blue) and recent theoretical predictions from NLEFT (purple) and IT-NCSM (green). The horizontal error bars on the NQS points represent combined Monte Carlo statistical uncertainties and uncertainties in the NN$\Lambda$ force.
   \label{fig:separation_energies_2}}
\end{figure}

Figure~\ref{fig:separation_energies_2} illustrates the $\Lambda$-separation energies of selected hypernuclei up to $^{16}_\Lambda$O. The NQS calculations are benchmarked against experimental values from the Hypernuclear Database~\cite{eckertChartHypernuclidesHypernuclear2023} (blue), as well as recent theoretical predictions~\cite{Knoll:2023mqk, Hildenbrand:2024ypw, Le:2024rkd} (green, purple, and red respectively), which differ both in the input Hamiltonians and in the numerical methods used to solve the nuclear Schr\"odinger equation. Specifically, Ref.~\cite{Knoll:2023mqk} employ the importance‑truncated no‑core shell model (IT–NCSM) using an N$^3$LO chiral‑EFT interaction for nucleons and an N$^3$LO N$\Lambda$ force, without explicit NN$\Lambda$ three‑body potentials. On the other hand, Ref.~\cite{Hildenbrand:2024ypw} use Nuclear Lattice Effective Field Theory (NLEFT) with N$^3$LO nucleon–nucleon interactions and leading‑order $s$‑wave N$\Lambda$ plus NN$\Lambda$ forces constrained by $A=4,5$ hypernuclei. Finally, Ref.~\cite{Le:2024rkd} report no‑core shell model (NCSM) calculations based on a consistent chiral expansion of both N$\Lambda$ and NN$\Lambda$ interactions up to N$^2$LO. 

The NQS energies are supplemented by estimates of their theoretical uncertainties, which include both the statistical error from the VMC calculation, and uncertainties associated with the NN$\Lambda$ potential: $\sigma_{B_\Lambda} =  \sigma_{\text{NQS}} + \sigma_{\Lambda\text{NN}}$. The latter are obtained in the same manner as in Figure~\ref{fig:energies_nuclei}, by considering ten different samples of the LECs and regulators entering the NN$\Lambda$ force, drawn from the joint posterior distribution shown in Figure~\ref{fig:correlation_lecs_hypernuclear}. For simplicity, the coupling and range of the NNN force are kept fixed to their median values, indicated by the orange dots in Figure~\ref{fig:correlation_lecs_nuclear}. This approximation is justified by the observation that the dependence of the $\Lambda$-separation energy on the nuclear Hamiltonian largely cancels when computing the energy difference between the hypernucleus and its parent nucleus~\cite{Lonardoni:2013rm}.

As expected, the NQS calculations for the systems used to calibrate the NN$\Lambda$ interaction closely reproduce the experimental data. The agreement remains excellent --- within theoretical and experimental uncertainties --- even for the $^4_\Lambda\text{He}_{S=1}$ and $^4_\Lambda\text{He}_{S=0}$ states, which differ from their $^4_\Lambda\text{H}$ analogues only due to small charge-dependent and charge-symmetry-breaking effects. Interestingly, the $B_\Lambda$ values of hypernuclei with $A \ge 6$ exhibit a nearly linear trend with the mass number $A$, up to $^{16}_\Lambda\text{O}$. Overall, the deviations from experimental data for hypernuclei not included in the fit remain small, with an average discrepancy below 9\%. The largest deviations --- significantly exceeding the combined theoretical and experimental uncertainties --- are observed for $^{12}_\Lambda\text{B}$, $^{12}_\Lambda\text{C}$, and $^{13}_\Lambda\text{C}$. This pattern is consistent with the underbinding observed in the corresponding parent nuclei and likely stems from the absence of $p$-wave contributions in both the NN and N$\Lambda$ potentials. We expect these relatively small discrepancies to be resolved upon inclusion of such contributions in these potentials.

\subsection{Hypernuclear Densities and Radii}
Spatial distribution functions and the corresponding radii provide additional insight into the structure of hypernuclei. In particular, they highlight the smaller N$\Lambda$ scattering length compared to NN interactions and illustrate the role of the Pauli exclusion principle in shaping nuclear structure. Notably, unlike quantum many-body approaches based on harmonic-oscillator basis expansions—such as the IT-NCSM and NCSM—the shell structure of nuclei and hypernuclei in NQS calculations emerges naturally from the variational minimization of the Hamiltonian expectation value, rather than being encoded in the variational ansatz.

The single-particle radial density distributions denote the probability of finding a nucleon or a $\Lambda$ particle at a distance $r$ from the center of mass of the hypernucleus. They are defined as the ground-state expectation value
\begin{equation}
\rho_\tau(r) = \frac{1}{4\pi r^2} \frac{\langle \Psi_{PJ}^{\Lambda} | \sum_i\delta(r - |\bar{\mathbf{r}}_i|) P^\tau_i | \Psi_{PJ}^{\Lambda} \rangle}{\langle \Psi_{PJ}^{\Lambda} | \Psi_{PJ}^{\Lambda} \rangle},  
\end{equation}
where $P^\tau_i$ is a proton, neutron, or $\Lambda$ projector. For illustrative purposes, we normalize the density distributions to unity, so that $\int 4\pi r^{2}\rho_\tau(r),dr=1$. In addition, to mitigate artifacts due to different spherical shell volumes, we multiply the density by a factor $r^2$.
The corresponding point radii are defined as
\begin{equation}
\langle r^2_\tau \rangle = \frac{1}{N_\tau} \frac{\langle \Psi_{PJ}^{\Lambda} | \sum_i \bar{\mathbf{r}}_i^2 P^\tau_i | \Psi_{PJ}^{\Lambda} \rangle}{\langle \Psi_{PJ}^{\Lambda} | \Psi_{PJ}^{\Lambda} \rangle},
\end{equation}
where $N_\tau$ denotes the number of protons or neutrons, and $N_\tau=1$ for the $\Lambda$ particle, since in this work we consider single-$\Lambda$ hypernuclei. 

To reduce statistical uncertainties, both the single-particle density distributions and the point radii are evaluated using approximately two million Monte Carlo configurations, binned into spherical shells of width $\Delta r = 0.04$fm. We note that, in contrast to diffusion Monte Carlo calculations, no extrapolations are required to compute expectation values of operators that do not commute with the Hamiltonian, as NQS states provide pure estimators for these quantities.

\begin{figure}[!htb]
\centering
    \hspace{-7pt}
    \includegraphics[width=0.45\textwidth]{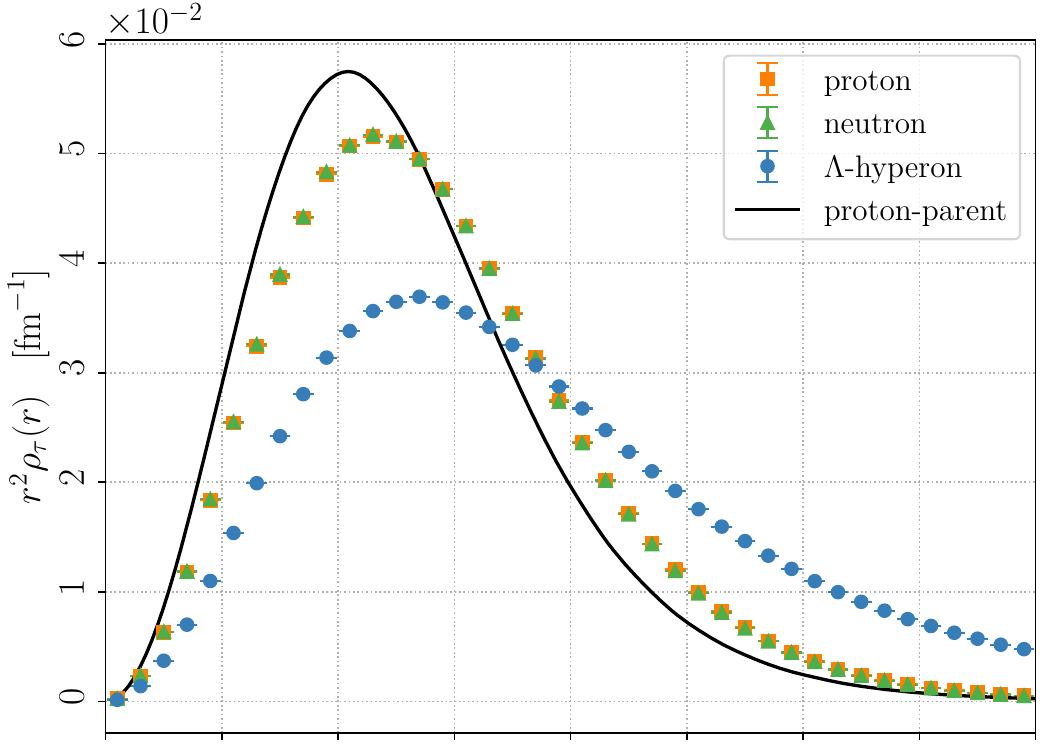}
    
    \hspace{-7pt}
    \includegraphics[width=0.45\textwidth]{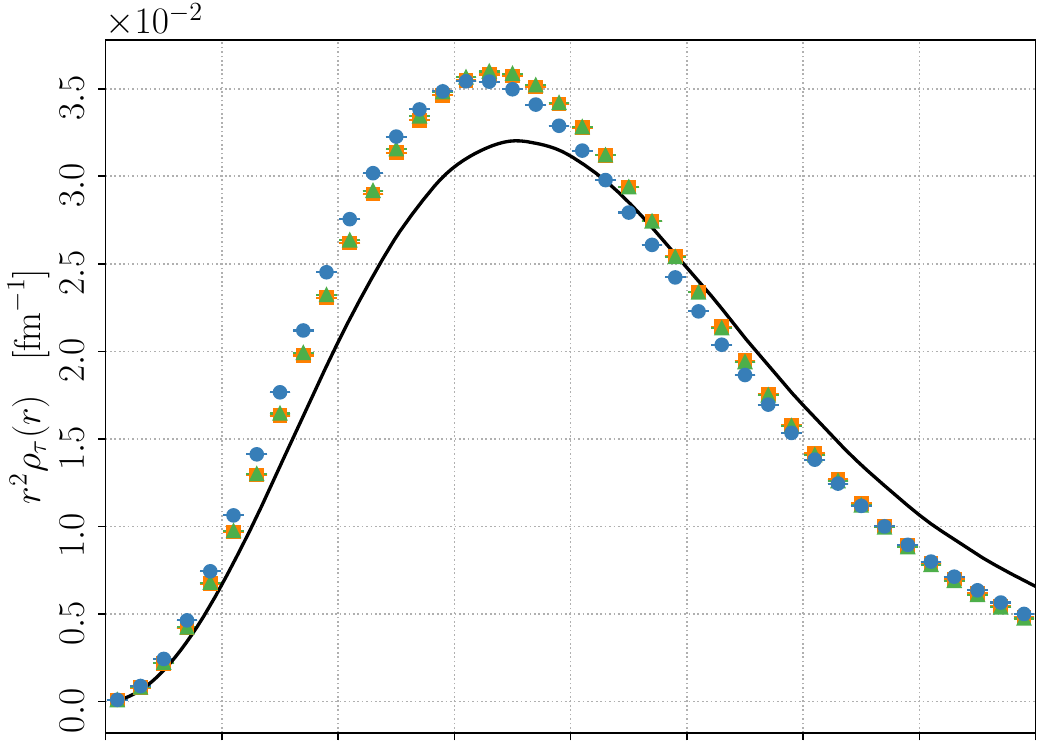}
    
    \includegraphics[width=0.46\textwidth]{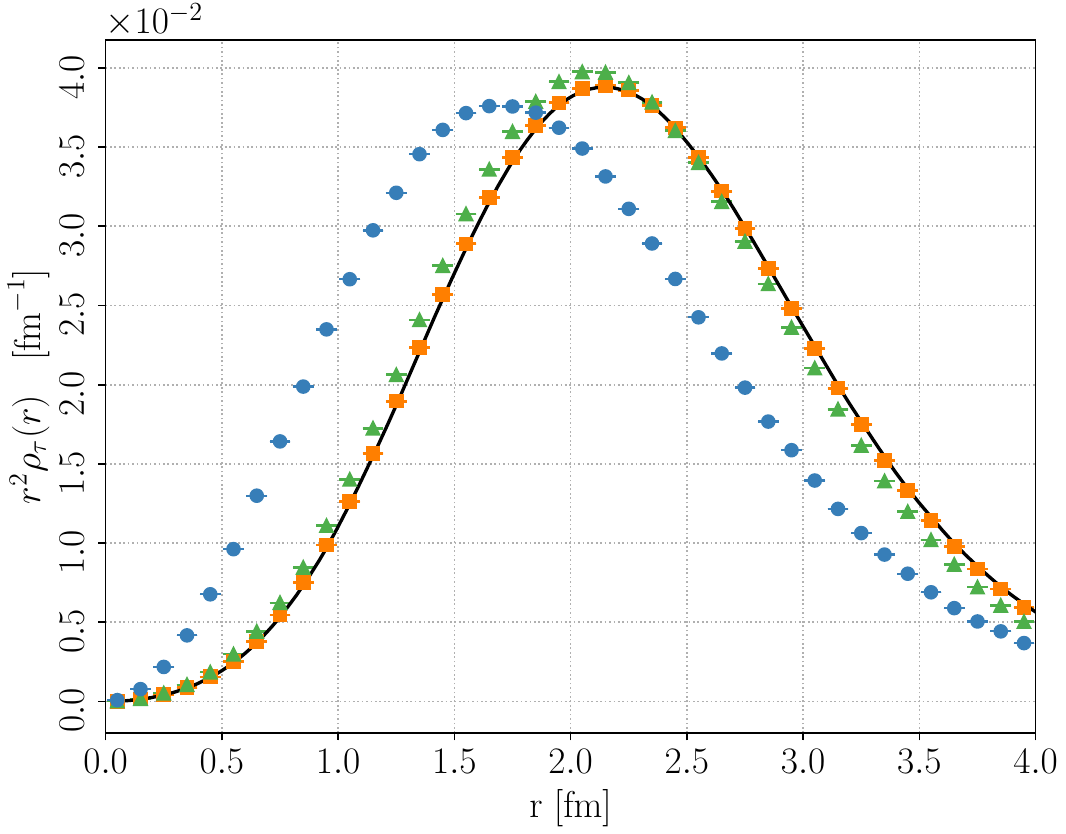}
    \caption{Neutron (orange squares), proton (green triangles), and $\Lambda$ hyperon (blue  circles) single-particle radial density distributions for $^5_\Lambda\textrm{He}$ (upper panel), $^7_\Lambda\textrm{Li}$ (middle panel), and $^{16}_\Lambda\textrm{O}$ (lower panel). The protons densities of the corresponding parent nuclei are also shown by the black solid line. For illustrative purposes, the distributions are normalized to unity and multiplied by a factor of $r^2$, as discussed in the text. \label{fig:rho_r}}
\end{figure}

The neutron, proton, and $\Lambda$-particle spatial density distributions of $^{5}_{\Lambda}\textrm{He}$, $^{7}_{\Lambda}\textrm{Li}$, and $^{16}_{\Lambda}\textrm{O}$ are displayed in the upper, middle, and lower panels of Figure~\ref{fig:rho_r}, respectively. In $^5_\Lambda\textrm{He}$, the $\Lambda$ hyperon is weakly bound and tends to form an extended halo around the $^4$He core. We observe similar behavior in other nuclei with $A < 7$, such as $^3_{\Lambda}$H, $^4_{\Lambda}$H, and $^4_{\Lambda}$He, indicating a significantly weaker attraction in the N$\Lambda$ potential compared to the NN interaction, consistent with the findings of Ref.~\cite{Hiyama:2001zt}. Starting from $^{7}_{\Lambda}$Li, the spatial distribution of the $\Lambda$ becomes more localized near the center of mass of the hypernucleus, compared to those of protons and neutrons. This behavior arises because the $\Lambda$, being distinguishable from nucleons, can continue to occupy inner $s$-shell orbitals, while protons and neutrons are subject to the Pauli exclusion principle and must populate higher $p$-shell states. This trend is further accentuated in $^{16}_{\Lambda}$O, where most protons and neutrons occupy $p$-shell states, while the $\Lambda$ remains free to reside in the central region of the system. 

\begin{table}[!t]
\centering
\renewcommand{\arraystretch}{1.25}
\caption{Point radii for protons, neutrons, and $\Lambda$ hyperon in $^5_\Lambda\textrm{He}$, $^7_\Lambda\textrm{Li}$, and $^{16}_\Lambda\textrm{O}$, along with those of their corresponding parent nuclei. All values are reported in fm, with statistical uncertainties shown in parentheses, affecting the last digit.}
\label{tab:radii}
\begin{tabular}{lccc}
\hline
System & $\sqrt{\langle \smash[b]{r^2_p} \rangle}$ & $\sqrt{\langle \smash[b]{r^2_n} \rangle}$ & $\sqrt{\langle \smash[b]{r^2_\Lambda} \rangle}$\\
\hline
$^5_{\Lambda}$He   & 1.57(1) & 1.57(1) & 2.32(1) \\
$^4$He             & 1.42(1) & 1.42(1)  & - \\
\hline
$^7_{\Lambda}$Li   & 2.30(1) & 2.30(1) & 2.30(2) \\
$^6\textrm{Li}$    & 2.64(2) & 2.63(2)  & -  \\
\hline
$^{16}_{\Lambda}$O & 2.48(1) & 2.41(1) & 2.15(1)  \\
$^{15}$O           & 2.50(1) & 2.42(1)  & -  \\ 
\hline
\end{tabular}
\end{table}

Further insights can be gained by comparing the corresponding radii of the three hypernuclei discussed above and their parent nuclei, as reported in Table~\ref{tab:radii}. Consistent with the density distributions, the radius of the $\Lambda$ hyperon is larger than those of the nucleons in $^5_\Lambda\textrm{He}$, becomes comparable in $^7_\Lambda\textrm{Li}$, and is significantly smaller in $^{16}_\Lambda$O. Even more revealing is the so-called ``core shrinkage,'' i.e., the relative change in the proton radius induced by the presence of the $\Lambda$ hyperon~\cite{Hiyama:1996gv,Hiyama:1999me}. To quantify this effect, we define the relative radius difference as
\begin{equation}
\delta r = \frac{\sqrt{\langle \smash[b]{r^2_p \rangle^A_\Lambda}} - \sqrt{\langle \smash[b]{r^2_p \rangle^{A-1}}}}{\sqrt{\langle \smash[b]{r^2_p \rangle^{A-1}}}}\,,
\end{equation}
where $\sqrt{\langle \smash[b]{r^2_p \rangle_{A}^{\Lambda}}}$ is the proton point radius of the hypernucleus, and $\sqrt{\langle \smash[b]{r^2_p} \rangle_{A-1}}$ is the corresponding radius for its parent nucleus. 

We observe a two-regime behavior of $\delta r$ as a function of $A$, which is consistent with trends observed in the spatial density distributions displayed in Figure~\ref{fig:rho_r}, where the proton density distribution of parent nuclei is given by a thick solid line. 

For light systems such as $^3_{\Lambda}$H, $^4_{\Lambda}$H, $^4_{\Lambda}$He, and $^5_{\Lambda}$He, our NQS calculations yield $\delta r > 0$, indicating that the proton radius in the hypernucleus is larger than in the parent nucleus. In these systems, the $\Lambda$ hyperon is weakly bound and tends to form an extended halo around the core, effectively ``pulling'' the nucleons outward from the center of mass. For $^5_\Lambda$He, we obtain $\delta r \simeq 0.11$, in line with the pioneering theoretical calculations of Ref.~\cite{Hiyama:2001zt} for $^4_\Lambda$H and $^4_\Lambda$He. 

Starting from $^7_\Lambda$Li --- for which we find $\delta r \simeq -0.13$ --- the proton radius of the hypernucleus becomes smaller than that of its parent nucleus, in qualitative agreement with the experimental findings of Ref.~\cite{Tanida:2000zs}. Within an extreme cluster model, $^6$Li can be interpreted as having an $\alpha$–deuteron clustered wave function, characterized by a relatively loose configuration and sizable separation between the clusters. When a $\Lambda$ hyperon is added to form $^7_\Lambda$Li, it occupies the central $1s$ orbital and interacts attractively with nucleons in both the $\alpha$ and $d$ clusters. Since the $\Lambda$ hyperon is distinguishable from nucleons and does not experience Pauli repulsion, it acts as a ``glue',' pulling the clusters closer together~\cite{Hiyama:1996gv}. This behavior is reflected in the central panel of Figure~\ref{fig:rho_r}, where the proton density of $^6$Li is broader than the one of $^7_\Lambda$Li.

We observe this glue-like effect gradually weakens with increasing $A$, decreasing to $\delta r \lesssim 0.01$ in $^{16}_\Lambda$O. This percent-level compaction agrees with the relativistic mean-field calculations of Ref.~\cite{Yao:2011wp}, which predict only slight hyperon-induced modifications to the core radius in this and heavier mass regions. Importantly, the fact that the $\Lambda$ hyperon resides precisely in the saturation-density region of the core enables direct calibration of the NN$\Lambda$ interaction in a density regime relevant to astrophysical applications~\cite{Lonardoni:2014bwa}.

\section{Conclusions}
We combine two complementary machine-learning approaches to compute ground-state properties of single-$\Lambda$ hypernuclei up to $^{16}_\Lambda$O, as they emerge from interacting neutrons, protons, and $\Lambda$ hyperons. Generalizing the paradigm introduced in Ref.~\cite{Lu:2018bat} to the strange sector, we seek the simplest Hamiltonian capable of reproducing the $\Lambda$ separation energies of $s$- and $p$-shell hypernuclei to within a few percent of their experimental values. Building on the arguments of Refs.~\cite{Schiavilla:2021dun} and \cite{Gnech:2023prs}, we consistently construct nuclear and hypernuclear two- and three-body interactions based on an improved leading-order pionless-EFT expansion. To avoid the instability problem of nuclei with more than $A=4$ nucleons, the range and couplings of the two-body NN and N$\Lambda$ potentials are fixed by reproducing the scattering length and effective range of $np$ and $p\Lambda$ scattering in the ${}^1S_0$ and ${}^3S_1$ channels~\cite{Schiavilla:2021dun,Contessi:2025xue}. 

Joint posterior distributions for the range and coupling constants entering both the three-body NNN and NN$\Lambda$ interactions are efficiently computed by leveraging Gaussian Processes to interpolate high-fidelity calculations of light nuclei and hypernuclei, performed using the SVM method~\cite{suzukiStochasticVariationalApproach1998c}. These posterior distributions directly reflect the relatively large experimental uncertainties of the measured $\Lambda$ separation energies. To improve the stability of the fitting procedure, the experimental errors on the nuclear binding energies are inflated to 2.5\% of their values. Our statistical analysis confirms the strong correlations among the ground-state energies of light nuclei; thus considering $^{16}$O is essential to constrain both the range and coupling of the NNN potential. A similar, though less pronounced, behavior is observed in light hypernuclei, indicating that the inclusion of the $^{16}_\Lambda$O binding energy is also necessary to effectively constrain the NN$\Lambda$ interaction. 

We solve the nuclear quantum many-body problem using a VMC method based on a highly flexible NQS ansatz, representing the first application of this approach to hypernuclear systems. Specifically, we generalize the Pfaffian-Jastrow architecture---originally introduced to model ultra-cold Fermi gases~\cite{Kim:2023fwy} and later extended to nuclear systems~\cite{Fore:2024exa}---to include hyperons alongside protons and neutrons. A key feature of our extension is that nuclear and hypernuclear forces do not mix $\Lambda$ hyperons and nucleons, allowing them to be treated as distinguishable particles. As a result, the coordinate of the $\Lambda$ hyperon does not need to be antisymmetrized with those of the nucleons. Additionally, we assign the $\Lambda$ an isospin value distinct from that of protons and neutrons, enabling both Jastrow and backflow correlations, represented by a MPNN~\cite{Pescia:2023mcc}, to effectively distinguish between $\Lambda$ hyperons, protons, and neutrons. In light hypernuclei, we succesfully benchmark this VNC-NQS approach against highly-accurate SVM calculations. 

Using the VMC-NQS method, we accurately compute ground-state properties of selected hypernuclei up to $^{16}_\Lambda$O. As a preliminary validation, we assess the accuracy of our interaction model by computing the binding energies of the parent nuclei, which we find to be within 5\% of the experimental values for systems not included in the determination of the NNN force. Similar agreement is obtained for the computed $\Lambda$ separation energies, with an average discrepancy of less than 9\% relative to experimental data. This larger deviation partly arises from simple error propagation, as $B_\Lambda$ is obtained by subtracting the ground-state energy of the hypernucleus from that of its parent nucleus—even though the dependence on the underlying nuclear Hamiltonian is expected to largely cancel in this energy difference. To estimate the uncertainties associated with the NN$\Lambda$ potential, we employ ten different parameterizations sampled from the joint posterior distribution. Our analysis suggests that discrepancies between theoretical predictions and experimental results likely originate from missing subleading terms in both nuclear and hypernuclear forces. In particular, we anticipate that including p-wave contributions~\cite{Gattobigio:2019omi} will help resolve the most significant deviations observed in $^{11}$B, $^{11}$C, and in $^{12}_\Lambda$B, $^{12}_\Lambda$C, and $^{13}_\Lambda$C.

Single-particle density distributions and radii provide further insights into the structure of hypernuclei, particularly highlighting the interplay between the relatively weak N$\Lambda$ attraction and the Pauli exclusion principle. In hypernuclei with $A \leq 5$, the $\Lambda$ hyperon is weakly bound and forms an extended halo around the parent nucleus, effectively pulling the nucleons outward and increasing the relative radius difference compared to the parent nucleus. Conversely, when a $\Lambda$ hyperon is introduced into a halo nucleus such as $^6$Li, it acts as a ``glue,'' reducing the relative radius difference, consistent with experimental observations~\cite{Tanida:2000zs} and earlier theoretical models~\cite{Hiyama:1996gv}. This glue-like effect diminishes with increasing nucleon number $A$, becoming nearly negligible in $^{16}_\Lambda$O. Importantly, in $^{16}_\Lambda$O, the $\Lambda$ primarily resides near the system's center of mass, occupying the central 1s orbital due to its distinguishability from nucleons, thereby enabling direct calibration of the NN$\Lambda$ interaction in a density regime relevant to astrophysical applications.

Given the favorable scalability of the VMC-NQS method with the number of interacting particles, the approach discussed here readily allows calculations for medium-mass hypernuclei such as $^{40}_\Lambda$Ca and $^{48}_\Lambda$Ca. Both of these nuclei, in addition to having been measured experimentally~\cite{eckertChartHypernuclidesHypernuclear2023}, exhibit extended regions with nuclear densities near saturation density $\rho_0\simeq 0.16$ fm$^{-3}$, close to the density found in neutron star cores. Furthermore, studying $^{40}_\Lambda$Ca and $^{48}_\Lambda$Ca will enable us to constrain potential isospin dependence of the NN$\Lambda$ interaction, as proposed in Ref.~\cite{Lonardoni:2017uuu}. To improve the baseline energies of the parent nuclei, we plan to include the leading p-wave contributions to the NN potential as in Ref.~\cite{Gattobigio:2019omi}.

Considering the high densities characteristic of neutron star interiors, systematic investigations of hyperon onset require using Hamiltonians with higher resolution than those employed in this work. Therefore, we plan to employ the local chiral-EFT NN and NNN interactions developed in Ref.~\cite{Piarulli:2017dwd}, potentially combining them with consistent N$\Lambda$ and NN$\Lambda$ interactions as in Ref.~\cite{Le:2024rkd}. The machine-learning methodology developed in this work will be particularly valuable, as it allows for efficient sensitivity analyses of chiral-EFT nuclear and hypernuclear forces and streamlined calculations of light- to medium-mass hypernuclei.

\section*{Acknowledgements}
The Authors acknowledge stimulating discussions with Alessandro Roggero. The present research is supported by the U.S. Department of Energy, Office of Science, Office of Nuclear Physics, under contracts DE-AC02-06CH11357 (A.~L.), by the DOE Early Career Research Program (A.~L.), by the STREAMLINE Collaboration Award No. DE-SC0024586 (A.~L., F.~P., and A.~D.~D.) and by the SciDAC-5 NUCLEI program (A.~L.). Numerical calculations were performed using an ISCRA award for accessing the LEONARDO supercomputer, owned by the EuroHPC Joint Undertaking, hosted by CINECA (Italy).

\bibliography{biblio.bib}

\begin{thebibliography}{87}%
\makeatletter
\providecommand \@ifxundefined [1]{%
 \@ifx{#1\undefined}
}%
\providecommand \@ifnum [1]{%
 \ifnum #1\expandafter \@firstoftwo
 \else \expandafter \@secondoftwo
 \fi
}%
\providecommand \@ifx [1]{%
 \ifx #1\expandafter \@firstoftwo
 \else \expandafter \@secondoftwo
 \fi
}%
\providecommand \natexlab [1]{#1}%
\providecommand \enquote  [1]{``#1''}%
\providecommand \bibnamefont  [1]{#1}%
\providecommand \bibfnamefont [1]{#1}%
\providecommand \citenamefont [1]{#1}%
\providecommand \href@noop [0]{\@secondoftwo}%
\providecommand \href [0]{\begingroup \@sanitize@url \@href}%
\providecommand \@href[1]{\@@startlink{#1}\@@href}%
\providecommand \@@href[1]{\endgroup#1\@@endlink}%
\providecommand \@sanitize@url [0]{\catcode `\\12\catcode `\$12\catcode
  `\&12\catcode `\#12\catcode `\^12\catcode `\_12\catcode `\%12\relax}%
\providecommand \@@startlink[1]{}%
\providecommand \@@endlink[0]{}%
\providecommand \url  [0]{\begingroup\@sanitize@url \@url }%
\providecommand \@url [1]{\endgroup\@href {#1}{\urlprefix }}%
\providecommand \urlprefix  [0]{URL }%
\providecommand \Eprint [0]{\href }%
\providecommand \doibase [0]{http://dx.doi.org/}%
\providecommand \selectlanguage [0]{\@gobble}%
\providecommand \bibinfo  [0]{\@secondoftwo}%
\providecommand \bibfield  [0]{\@secondoftwo}%
\providecommand \translation [1]{[#1]}%
\providecommand \BibitemOpen [0]{}%
\providecommand \bibitemStop [0]{}%
\providecommand \bibitemNoStop [0]{.\EOS\space}%
\providecommand \EOS [0]{\spacefactor3000\relax}%
\providecommand \BibitemShut  [1]{\csname bibitem#1\endcsname}%
\let\auto@bib@innerbib\@empty
\bibitem [{\citenamefont {Epelbaum}\ \emph {et~al.}(2009)\citenamefont
  {Epelbaum}, \citenamefont {Hammer},\ and\ \citenamefont
  {Meissner}}]{Epelbaum:2008ga}%
  \BibitemOpen
  \bibfield  {author} {\bibinfo {author} {\bibfnamefont {E.}~\bibnamefont
  {Epelbaum}}, \bibinfo {author} {\bibfnamefont {H.-W.}\ \bibnamefont
  {Hammer}}, \ and\ \bibinfo {author} {\bibfnamefont {U.-G.}\ \bibnamefont
  {Meissner}},\ }\href {\doibase 10.1103/RevModPhys.81.1773} {\bibfield
  {journal} {\bibinfo  {journal} {Rev. Mod. Phys.}\ }\textbf {\bibinfo {volume}
  {81}},\ \bibinfo {pages} {1773} (\bibinfo {year} {2009})},\ \Eprint
  {http://arxiv.org/abs/0811.1338} {arXiv:0811.1338 [nucl-th]} \BibitemShut
  {NoStop}%
\bibitem [{\citenamefont {Machleidt}\ and\ \citenamefont
  {Entem}(2011)}]{Machleidt:2011zz}%
  \BibitemOpen
  \bibfield  {author} {\bibinfo {author} {\bibfnamefont {R.}~\bibnamefont
  {Machleidt}}\ and\ \bibinfo {author} {\bibfnamefont {D.~R.}\ \bibnamefont
  {Entem}},\ }\href {\doibase 10.1016/j.physrep.2011.02.001} {\bibfield
  {journal} {\bibinfo  {journal} {Phys. Rept.}\ }\textbf {\bibinfo {volume}
  {503}},\ \bibinfo {pages} {1} (\bibinfo {year} {2011})},\ \Eprint
  {http://arxiv.org/abs/1105.2919} {arXiv:1105.2919 [nucl-th]} \BibitemShut
  {NoStop}%
\bibitem [{\citenamefont {Hammer}\ \emph {et~al.}(2020)\citenamefont {Hammer},
  \citenamefont {K\"onig},\ and\ \citenamefont {van Kolck}}]{Hammer:2019poc}%
  \BibitemOpen
  \bibfield  {author} {\bibinfo {author} {\bibfnamefont {H.~W.}\ \bibnamefont
  {Hammer}}, \bibinfo {author} {\bibfnamefont {S.}~\bibnamefont {K\"onig}}, \
  and\ \bibinfo {author} {\bibfnamefont {U.}~\bibnamefont {van Kolck}},\ }\href
  {\doibase 10.1103/RevModPhys.92.025004} {\bibfield  {journal} {\bibinfo
  {journal} {Rev. Mod. Phys.}\ }\textbf {\bibinfo {volume} {92}},\ \bibinfo
  {pages} {025004} (\bibinfo {year} {2020})},\ \Eprint
  {http://arxiv.org/abs/1906.12122} {arXiv:1906.12122 [nucl-th]} \BibitemShut
  {NoStop}%
\bibitem [{\citenamefont {Svensson}\ \emph {et~al.}(2022)\citenamefont
  {Svensson}, \citenamefont {Ekstr\"om},\ and\ \citenamefont
  {Forss\'en}}]{Svensson:2021lzs}%
  \BibitemOpen
  \bibfield  {author} {\bibinfo {author} {\bibfnamefont {I.}~\bibnamefont
  {Svensson}}, \bibinfo {author} {\bibfnamefont {A.}~\bibnamefont {Ekstr\"om}},
  \ and\ \bibinfo {author} {\bibfnamefont {C.}~\bibnamefont {Forss\'en}},\
  }\href {\doibase 10.1103/PhysRevC.105.014004} {\bibfield  {journal} {\bibinfo
   {journal} {Phys. Rev. C}\ }\textbf {\bibinfo {volume} {105}},\ \bibinfo
  {pages} {014004} (\bibinfo {year} {2022})},\ \Eprint
  {http://arxiv.org/abs/2110.04011} {arXiv:2110.04011 [nucl-th]} \BibitemShut
  {NoStop}%
\bibitem [{\citenamefont {Bub}\ \emph {et~al.}(2025)\citenamefont {Bub},
  \citenamefont {Piarulli}, \citenamefont {Furnstahl}, \citenamefont
  {Pastore},\ and\ \citenamefont {Phillips}}]{Bub:2024gyz}%
  \BibitemOpen
  \bibfield  {author} {\bibinfo {author} {\bibfnamefont {J.~M.}\ \bibnamefont
  {Bub}}, \bibinfo {author} {\bibfnamefont {M.}~\bibnamefont {Piarulli}},
  \bibinfo {author} {\bibfnamefont {R.~J.}\ \bibnamefont {Furnstahl}}, \bibinfo
  {author} {\bibfnamefont {S.}~\bibnamefont {Pastore}}, \ and\ \bibinfo
  {author} {\bibfnamefont {D.~R.}\ \bibnamefont {Phillips}},\ }\href {\doibase
  10.1103/PhysRevC.111.034005} {\bibfield  {journal} {\bibinfo  {journal}
  {Phys. Rev. C}\ }\textbf {\bibinfo {volume} {111}},\ \bibinfo {pages}
  {034005} (\bibinfo {year} {2025})},\ \Eprint
  {http://arxiv.org/abs/2408.02480} {arXiv:2408.02480 [nucl-th]} \BibitemShut
  {NoStop}%
\bibitem [{\citenamefont {Hagen}\ \emph {et~al.}(2014)\citenamefont {Hagen},
  \citenamefont {Papenbrock}, \citenamefont {Hjorth-Jensen},\ and\
  \citenamefont {Dean}}]{Hagen:2013nca}%
  \BibitemOpen
  \bibfield  {author} {\bibinfo {author} {\bibfnamefont {G.}~\bibnamefont
  {Hagen}}, \bibinfo {author} {\bibfnamefont {T.}~\bibnamefont {Papenbrock}},
  \bibinfo {author} {\bibfnamefont {M.}~\bibnamefont {Hjorth-Jensen}}, \ and\
  \bibinfo {author} {\bibfnamefont {D.~J.}\ \bibnamefont {Dean}},\ }\href
  {\doibase 10.1088/0034-4885/77/9/096302} {\bibfield  {journal} {\bibinfo
  {journal} {Rept. Prog. Phys.}\ }\textbf {\bibinfo {volume} {77}},\ \bibinfo
  {pages} {096302} (\bibinfo {year} {2014})},\ \Eprint
  {http://arxiv.org/abs/1312.7872} {arXiv:1312.7872 [nucl-th]} \BibitemShut
  {NoStop}%
\bibitem [{\citenamefont {Hergert}\ \emph {et~al.}(2016)\citenamefont
  {Hergert}, \citenamefont {Bogner}, \citenamefont {Morris}, \citenamefont
  {Schwenk},\ and\ \citenamefont {Tsukiyama}}]{Hergert:2015awm}%
  \BibitemOpen
  \bibfield  {author} {\bibinfo {author} {\bibfnamefont {H.}~\bibnamefont
  {Hergert}}, \bibinfo {author} {\bibfnamefont {S.~K.}\ \bibnamefont {Bogner}},
  \bibinfo {author} {\bibfnamefont {T.~D.}\ \bibnamefont {Morris}}, \bibinfo
  {author} {\bibfnamefont {A.}~\bibnamefont {Schwenk}}, \ and\ \bibinfo
  {author} {\bibfnamefont {K.}~\bibnamefont {Tsukiyama}},\ }\href {\doibase
  10.1016/j.physrep.2015.12.007} {\bibfield  {journal} {\bibinfo  {journal}
  {Phys. Rept.}\ }\textbf {\bibinfo {volume} {621}},\ \bibinfo {pages} {165}
  (\bibinfo {year} {2016})},\ \Eprint {http://arxiv.org/abs/1512.06956}
  {arXiv:1512.06956 [nucl-th]} \BibitemShut {NoStop}%
\bibitem [{\citenamefont {Dickhoff}\ and\ \citenamefont
  {Barbieri}(2004)}]{Dickhoff:2004xx}%
  \BibitemOpen
  \bibfield  {author} {\bibinfo {author} {\bibfnamefont {W.~H.}\ \bibnamefont
  {Dickhoff}}\ and\ \bibinfo {author} {\bibfnamefont {C.}~\bibnamefont
  {Barbieri}},\ }\href {\doibase 10.1016/j.ppnp.2004.02.038} {\bibfield
  {journal} {\bibinfo  {journal} {Prog. Part. Nucl. Phys.}\ }\textbf {\bibinfo
  {volume} {52}},\ \bibinfo {pages} {377} (\bibinfo {year} {2004})},\ \Eprint
  {http://arxiv.org/abs/nucl-th/0402034} {arXiv:nucl-th/0402034} \BibitemShut
  {NoStop}%
\bibitem [{\citenamefont {Soma}\ \emph {et~al.}(2013)\citenamefont {Soma},
  \citenamefont {Barbieri},\ and\ \citenamefont {Duguet}}]{Soma:2012zd}%
  \BibitemOpen
  \bibfield  {author} {\bibinfo {author} {\bibfnamefont {V.}~\bibnamefont
  {Soma}}, \bibinfo {author} {\bibfnamefont {C.}~\bibnamefont {Barbieri}}, \
  and\ \bibinfo {author} {\bibfnamefont {T.}~\bibnamefont {Duguet}},\ }\href
  {\doibase 10.1103/PhysRevC.87.011303} {\bibfield  {journal} {\bibinfo
  {journal} {Phys. Rev. C}\ }\textbf {\bibinfo {volume} {87}},\ \bibinfo
  {pages} {011303} (\bibinfo {year} {2013})},\ \Eprint
  {http://arxiv.org/abs/1208.2472} {arXiv:1208.2472 [nucl-th]} \BibitemShut
  {NoStop}%
\bibitem [{\citenamefont {Carlson}\ \emph {et~al.}(2015)\citenamefont
  {Carlson}, \citenamefont {Gandolfi}, \citenamefont {Pederiva}, \citenamefont
  {Pieper}, \citenamefont {Schiavilla}, \citenamefont {Schmidt},\ and\
  \citenamefont {Wiringa}}]{Carlson:2014vla}%
  \BibitemOpen
  \bibfield  {author} {\bibinfo {author} {\bibfnamefont {J.}~\bibnamefont
  {Carlson}}, \bibinfo {author} {\bibfnamefont {S.}~\bibnamefont {Gandolfi}},
  \bibinfo {author} {\bibfnamefont {F.}~\bibnamefont {Pederiva}}, \bibinfo
  {author} {\bibfnamefont {S.~C.}\ \bibnamefont {Pieper}}, \bibinfo {author}
  {\bibfnamefont {R.}~\bibnamefont {Schiavilla}}, \bibinfo {author}
  {\bibfnamefont {K.~E.}\ \bibnamefont {Schmidt}}, \ and\ \bibinfo {author}
  {\bibfnamefont {R.~B.}\ \bibnamefont {Wiringa}},\ }\href {\doibase
  10.1103/RevModPhys.87.1067} {\bibfield  {journal} {\bibinfo  {journal} {Rev.
  Mod. Phys.}\ }\textbf {\bibinfo {volume} {87}},\ \bibinfo {pages} {1067}
  (\bibinfo {year} {2015})},\ \Eprint {http://arxiv.org/abs/1412.3081}
  {arXiv:1412.3081 [nucl-th]} \BibitemShut {NoStop}%
\bibitem [{\citenamefont {Lynn}\ \emph {et~al.}(2019)\citenamefont {Lynn},
  \citenamefont {Tews}, \citenamefont {Gandolfi},\ and\ \citenamefont
  {Lovato}}]{Lynn:2019rdt}%
  \BibitemOpen
  \bibfield  {author} {\bibinfo {author} {\bibfnamefont {J.~E.}\ \bibnamefont
  {Lynn}}, \bibinfo {author} {\bibfnamefont {I.}~\bibnamefont {Tews}}, \bibinfo
  {author} {\bibfnamefont {S.}~\bibnamefont {Gandolfi}}, \ and\ \bibinfo
  {author} {\bibfnamefont {A.}~\bibnamefont {Lovato}},\ }\href {\doibase
  10.1146/annurev-nucl-101918-023600} {\bibfield  {journal} {\bibinfo
  {journal} {Ann. Rev. Nucl. Part. Sci.}\ }\textbf {\bibinfo {volume} {69}},\
  \bibinfo {pages} {279} (\bibinfo {year} {2019})},\ \Eprint
  {http://arxiv.org/abs/1901.04868} {arXiv:1901.04868 [nucl-th]} \BibitemShut
  {NoStop}%
\bibitem [{\citenamefont {Danysz}\ and\ \citenamefont
  {Pniewski}(1953)}]{Danysz:1953}%
  \BibitemOpen
  \bibfield  {author} {\bibinfo {author} {\bibfnamefont {M.}~\bibnamefont
  {Danysz}}\ and\ \bibinfo {author} {\bibfnamefont {J.}~\bibnamefont
  {Pniewski}},\ }\href {\doibase 10.1080/14786440308520318} {\bibfield
  {journal} {\bibinfo  {journal} {The London, Edinburgh, and Dublin
  Philosophical Magazine and Journal of Science}\ }\textbf {\bibinfo {volume}
  {44}},\ \bibinfo {pages} {348} (\bibinfo {year} {1953})},\ \Eprint
  {http://arxiv.org/abs/https://doi.org/10.1080/14786440308520318}
  {https://doi.org/10.1080/14786440308520318} \BibitemShut {NoStop}%
\bibitem [{\citenamefont {May}\ \emph {et~al.}(1983)\citenamefont {May} \emph
  {et~al.}}]{May:1983oof}%
  \BibitemOpen
  \bibfield  {author} {\bibinfo {author} {\bibfnamefont {M.}~\bibnamefont
  {May}} \emph {et~al.},\ }\href {\doibase 10.1103/PhysRevLett.51.2085}
  {\bibfield  {journal} {\bibinfo  {journal} {Phys. Rev. Lett.}\ }\textbf
  {\bibinfo {volume} {51}},\ \bibinfo {pages} {2085} (\bibinfo {year}
  {1983})}\BibitemShut {NoStop}%
\bibitem [{\citenamefont {Tanida}\ \emph {et~al.}(2001)\citenamefont {Tanida}
  \emph {et~al.}}]{Tanida:2000zs}%
  \BibitemOpen
  \bibfield  {author} {\bibinfo {author} {\bibfnamefont {K.}~\bibnamefont
  {Tanida}} \emph {et~al.},\ }\href {\doibase 10.1103/PhysRevLett.86.1982}
  {\bibfield  {journal} {\bibinfo  {journal} {Phys. Rev. Lett.}\ }\textbf
  {\bibinfo {volume} {86}},\ \bibinfo {pages} {1982} (\bibinfo {year}
  {2001})}\BibitemShut {NoStop}%
\bibitem [{\citenamefont {Akikawa}\ \emph {et~al.}(2002)\citenamefont {Akikawa}
  \emph {et~al.}}]{Akikawa:2002tm}%
  \BibitemOpen
  \bibfield  {author} {\bibinfo {author} {\bibfnamefont {H.}~\bibnamefont
  {Akikawa}} \emph {et~al.},\ }\href {\doibase 10.1103/PhysRevLett.88.082501}
  {\bibfield  {journal} {\bibinfo  {journal} {Phys. Rev. Lett.}\ }\textbf
  {\bibinfo {volume} {88}},\ \bibinfo {pages} {082501} (\bibinfo {year}
  {2002})}\BibitemShut {NoStop}%
\bibitem [{\citenamefont {Hashimoto}\ and\ \citenamefont
  {Tamura}(2006)}]{Hashimoto:2006aw}%
  \BibitemOpen
  \bibfield  {author} {\bibinfo {author} {\bibfnamefont {O.}~\bibnamefont
  {Hashimoto}}\ and\ \bibinfo {author} {\bibfnamefont {H.}~\bibnamefont
  {Tamura}},\ }\href {\doibase 10.1016/j.ppnp.2005.07.001} {\bibfield
  {journal} {\bibinfo  {journal} {Prog. Part. Nucl. Phys.}\ }\textbf {\bibinfo
  {volume} {57}},\ \bibinfo {pages} {564} (\bibinfo {year} {2006})}\BibitemShut
  {NoStop}%
\bibitem [{\citenamefont {Yamamoto}\ \emph {et~al.}(2015)\citenamefont
  {Yamamoto} \emph {et~al.}}]{J-PARCE13:2015uwb}%
  \BibitemOpen
  \bibfield  {author} {\bibinfo {author} {\bibfnamefont {T.~O.}\ \bibnamefont
  {Yamamoto}} \emph {et~al.} (\bibinfo {collaboration} {J-PARC E13}),\ }\href
  {\doibase 10.1103/PhysRevLett.115.222501} {\bibfield  {journal} {\bibinfo
  {journal} {Phys. Rev. Lett.}\ }\textbf {\bibinfo {volume} {115}},\ \bibinfo
  {pages} {222501} (\bibinfo {year} {2015})},\ \Eprint
  {http://arxiv.org/abs/1508.00376} {arXiv:1508.00376 [nucl-ex]} \BibitemShut
  {NoStop}%
\bibitem [{\citenamefont {Gogami}\ \emph {et~al.}(2016)\citenamefont {Gogami}
  \emph {et~al.}}]{Gogami:2015tvu}%
  \BibitemOpen
  \bibfield  {author} {\bibinfo {author} {\bibfnamefont {T.}~\bibnamefont
  {Gogami}} \emph {et~al.},\ }\href {\doibase 10.1103/PhysRevC.93.034314}
  {\bibfield  {journal} {\bibinfo  {journal} {Phys. Rev. C}\ }\textbf {\bibinfo
  {volume} {93}},\ \bibinfo {pages} {034314} (\bibinfo {year} {2016})},\
  \Eprint {http://arxiv.org/abs/1511.04801} {arXiv:1511.04801 [nucl-ex]}
  \BibitemShut {NoStop}%
\bibitem [{\citenamefont {Gal}\ \emph {et~al.}(2016)\citenamefont {Gal},
  \citenamefont {Hungerford},\ and\ \citenamefont {Millener}}]{Gal:2016boi}%
  \BibitemOpen
  \bibfield  {author} {\bibinfo {author} {\bibfnamefont {A.}~\bibnamefont
  {Gal}}, \bibinfo {author} {\bibfnamefont {E.~V.}\ \bibnamefont {Hungerford}},
  \ and\ \bibinfo {author} {\bibfnamefont {D.~J.}\ \bibnamefont {Millener}},\
  }\href {\doibase 10.1103/RevModPhys.88.035004} {\bibfield  {journal}
  {\bibinfo  {journal} {Rev. Mod. Phys.}\ }\textbf {\bibinfo {volume} {88}},\
  \bibinfo {pages} {035004} (\bibinfo {year} {2016})},\ \Eprint
  {http://arxiv.org/abs/1605.00557} {arXiv:1605.00557 [nucl-th]} \BibitemShut
  {NoStop}%
\bibitem [{\citenamefont {Bombaci}(2017)}]{Bombaci:2016xzl}%
  \BibitemOpen
  \bibfield  {author} {\bibinfo {author} {\bibfnamefont {I.}~\bibnamefont
  {Bombaci}},\ }\href {\doibase 10.7566/JPSCP.17.101002} {\bibfield  {journal}
  {\bibinfo  {journal} {JPS Conf. Proc.}\ }\textbf {\bibinfo {volume} {17}},\
  \bibinfo {pages} {101002} (\bibinfo {year} {2017})},\ \Eprint
  {http://arxiv.org/abs/1601.05339} {arXiv:1601.05339 [nucl-th]} \BibitemShut
  {NoStop}%
\bibitem [{\citenamefont {Maslov}\ \emph {et~al.}(2015)\citenamefont {Maslov},
  \citenamefont {Kolomeitsev},\ and\ \citenamefont
  {Voskresensky}}]{Maslov:2015msa}%
  \BibitemOpen
  \bibfield  {author} {\bibinfo {author} {\bibfnamefont {K.~A.}\ \bibnamefont
  {Maslov}}, \bibinfo {author} {\bibfnamefont {E.~E.}\ \bibnamefont
  {Kolomeitsev}}, \ and\ \bibinfo {author} {\bibfnamefont {D.~N.}\ \bibnamefont
  {Voskresensky}},\ }\href {\doibase 10.1016/j.physletb.2015.07.032} {\bibfield
   {journal} {\bibinfo  {journal} {Phys. Lett. B}\ }\textbf {\bibinfo {volume}
  {748}},\ \bibinfo {pages} {369} (\bibinfo {year} {2015})},\ \Eprint
  {http://arxiv.org/abs/1504.02915} {arXiv:1504.02915 [astro-ph.HE]}
  \BibitemShut {NoStop}%
\bibitem [{\citenamefont {Demorest}\ \emph {et~al.}(2010)\citenamefont
  {Demorest}, \citenamefont {Pennucci}, \citenamefont {Ransom}, \citenamefont
  {Roberts},\ and\ \citenamefont {Hessels}}]{Demorest:2010bx}%
  \BibitemOpen
  \bibfield  {author} {\bibinfo {author} {\bibfnamefont {P.}~\bibnamefont
  {Demorest}}, \bibinfo {author} {\bibfnamefont {T.}~\bibnamefont {Pennucci}},
  \bibinfo {author} {\bibfnamefont {S.}~\bibnamefont {Ransom}}, \bibinfo
  {author} {\bibfnamefont {M.}~\bibnamefont {Roberts}}, \ and\ \bibinfo
  {author} {\bibfnamefont {J.}~\bibnamefont {Hessels}},\ }\href {\doibase
  10.1038/nature09466} {\bibfield  {journal} {\bibinfo  {journal} {Nature}\
  }\textbf {\bibinfo {volume} {467}},\ \bibinfo {pages} {1081} (\bibinfo {year}
  {2010})},\ \Eprint {http://arxiv.org/abs/1010.5788} {arXiv:1010.5788
  [astro-ph.HE]} \BibitemShut {NoStop}%
\bibitem [{\citenamefont {Antoniadis}\ \emph {et~al.}(2013)\citenamefont
  {Antoniadis} \emph {et~al.}}]{Antoniadis:2013pzd}%
  \BibitemOpen
  \bibfield  {author} {\bibinfo {author} {\bibfnamefont {J.}~\bibnamefont
  {Antoniadis}} \emph {et~al.},\ }\href {\doibase 10.1126/science.1233232}
  {\bibfield  {journal} {\bibinfo  {journal} {Science}\ }\textbf {\bibinfo
  {volume} {340}},\ \bibinfo {pages} {6131} (\bibinfo {year} {2013})},\ \Eprint
  {http://arxiv.org/abs/1304.6875} {arXiv:1304.6875 [astro-ph.HE]} \BibitemShut
  {NoStop}%
\bibitem [{\citenamefont {Cromartie}\ \emph {et~al.}(2019)\citenamefont
  {Cromartie} \emph {et~al.}}]{NANOGrav:2019jur}%
  \BibitemOpen
  \bibfield  {author} {\bibinfo {author} {\bibfnamefont {H.~T.}\ \bibnamefont
  {Cromartie}} \emph {et~al.} (\bibinfo {collaboration} {NANOGrav}),\ }\href
  {\doibase 10.1038/s41550-019-0880-2} {\bibfield  {journal} {\bibinfo
  {journal} {Nature Astron.}\ }\textbf {\bibinfo {volume} {4}},\ \bibinfo
  {pages} {72} (\bibinfo {year} {2019})},\ \Eprint
  {http://arxiv.org/abs/1904.06759} {arXiv:1904.06759 [astro-ph.HE]}
  \BibitemShut {NoStop}%
\bibitem [{\citenamefont {Fonseca}\ \emph {et~al.}(2021)\citenamefont {Fonseca}
  \emph {et~al.}}]{Fonseca:2021wxt}%
  \BibitemOpen
  \bibfield  {author} {\bibinfo {author} {\bibfnamefont {E.}~\bibnamefont
  {Fonseca}} \emph {et~al.},\ }\href {\doibase 10.3847/2041-8213/ac03b8}
  {\bibfield  {journal} {\bibinfo  {journal} {Astrophys. J. Lett.}\ }\textbf
  {\bibinfo {volume} {915}},\ \bibinfo {pages} {L12} (\bibinfo {year}
  {2021})},\ \Eprint {http://arxiv.org/abs/2104.00880} {arXiv:2104.00880
  [astro-ph.HE]} \BibitemShut {NoStop}%
\bibitem [{\citenamefont {Lonardoni}\ \emph {et~al.}(2015)\citenamefont
  {Lonardoni}, \citenamefont {Lovato}, \citenamefont {Gandolfi},\ and\
  \citenamefont {Pederiva}}]{Lonardoni:2014bwa}%
  \BibitemOpen
  \bibfield  {author} {\bibinfo {author} {\bibfnamefont {D.}~\bibnamefont
  {Lonardoni}}, \bibinfo {author} {\bibfnamefont {A.}~\bibnamefont {Lovato}},
  \bibinfo {author} {\bibfnamefont {S.}~\bibnamefont {Gandolfi}}, \ and\
  \bibinfo {author} {\bibfnamefont {F.}~\bibnamefont {Pederiva}},\ }\href
  {\doibase 10.1103/PhysRevLett.114.092301} {\bibfield  {journal} {\bibinfo
  {journal} {Phys. Rev. Lett.}\ }\textbf {\bibinfo {volume} {114}},\ \bibinfo
  {pages} {092301} (\bibinfo {year} {2015})},\ \Eprint
  {http://arxiv.org/abs/1407.4448} {arXiv:1407.4448 [nucl-th]} \BibitemShut
  {NoStop}%
\bibitem [{\citenamefont {Haidenbauer}\ \emph {et~al.}(2017)\citenamefont
  {Haidenbauer}, \citenamefont {Mei\ss{}ner}, \citenamefont {Kaiser},\ and\
  \citenamefont {Weise}}]{Haidenbauer:2016vfq}%
  \BibitemOpen
  \bibfield  {author} {\bibinfo {author} {\bibfnamefont {J.}~\bibnamefont
  {Haidenbauer}}, \bibinfo {author} {\bibfnamefont {U.~G.}\ \bibnamefont
  {Mei\ss{}ner}}, \bibinfo {author} {\bibfnamefont {N.}~\bibnamefont {Kaiser}},
  \ and\ \bibinfo {author} {\bibfnamefont {W.}~\bibnamefont {Weise}},\ }\href
  {\doibase 10.1140/epja/i2017-12316-4} {\bibfield  {journal} {\bibinfo
  {journal} {Eur. Phys. J. A}\ }\textbf {\bibinfo {volume} {53}},\ \bibinfo
  {pages} {121} (\bibinfo {year} {2017})},\ \Eprint
  {http://arxiv.org/abs/1612.03758} {arXiv:1612.03758 [nucl-th]} \BibitemShut
  {NoStop}%
\bibitem [{\citenamefont {Nogga}\ \emph {et~al.}(2002)\citenamefont {Nogga},
  \citenamefont {Kamada},\ and\ \citenamefont {Gloeckle}}]{Nogga:2001ef}%
  \BibitemOpen
  \bibfield  {author} {\bibinfo {author} {\bibfnamefont {A.}~\bibnamefont
  {Nogga}}, \bibinfo {author} {\bibfnamefont {H.}~\bibnamefont {Kamada}}, \
  and\ \bibinfo {author} {\bibfnamefont {W.}~\bibnamefont {Gloeckle}},\ }\href
  {\doibase 10.1103/PhysRevLett.88.172501} {\bibfield  {journal} {\bibinfo
  {journal} {Phys. Rev. Lett.}\ }\textbf {\bibinfo {volume} {88}},\ \bibinfo
  {pages} {172501} (\bibinfo {year} {2002})},\ \Eprint
  {http://arxiv.org/abs/nucl-th/0112060} {arXiv:nucl-th/0112060} \BibitemShut
  {NoStop}%
\bibitem [{\citenamefont {Hiyama}\ \emph {et~al.}(2002)\citenamefont {Hiyama},
  \citenamefont {Kamimura}, \citenamefont {Motoba}, \citenamefont {Yamada},\
  and\ \citenamefont {Yamamoto}}]{Hiyama:2001zt}%
  \BibitemOpen
  \bibfield  {author} {\bibinfo {author} {\bibfnamefont {E.}~\bibnamefont
  {Hiyama}}, \bibinfo {author} {\bibfnamefont {M.}~\bibnamefont {Kamimura}},
  \bibinfo {author} {\bibfnamefont {T.}~\bibnamefont {Motoba}}, \bibinfo
  {author} {\bibfnamefont {T.}~\bibnamefont {Yamada}}, \ and\ \bibinfo {author}
  {\bibfnamefont {Y.}~\bibnamefont {Yamamoto}},\ }\href {\doibase
  10.1103/PhysRevC.65.011301} {\bibfield  {journal} {\bibinfo  {journal} {Phys.
  Rev. C}\ }\textbf {\bibinfo {volume} {65}},\ \bibinfo {pages} {011301}
  (\bibinfo {year} {2002})},\ \Eprint {http://arxiv.org/abs/nucl-th/0106070}
  {arXiv:nucl-th/0106070} \BibitemShut {NoStop}%
\bibitem [{\citenamefont {Nemura}\ \emph {et~al.}(2002)\citenamefont {Nemura},
  \citenamefont {Akaishi},\ and\ \citenamefont {Suzuki}}]{Nemura:2002fu}%
  \BibitemOpen
  \bibfield  {author} {\bibinfo {author} {\bibfnamefont {H.}~\bibnamefont
  {Nemura}}, \bibinfo {author} {\bibfnamefont {Y.}~\bibnamefont {Akaishi}}, \
  and\ \bibinfo {author} {\bibfnamefont {Y.}~\bibnamefont {Suzuki}},\ }\href
  {\doibase 10.1103/PhysRevLett.89.142504} {\bibfield  {journal} {\bibinfo
  {journal} {Phys. Rev. Lett.}\ }\textbf {\bibinfo {volume} {89}},\ \bibinfo
  {pages} {142504} (\bibinfo {year} {2002})},\ \Eprint
  {http://arxiv.org/abs/nucl-th/0203013} {arXiv:nucl-th/0203013} \BibitemShut
  {NoStop}%
\bibitem [{\citenamefont {Filikhin}\ and\ \citenamefont
  {Gal}(2002)}]{Filikhin:2002wp}%
  \BibitemOpen
  \bibfield  {author} {\bibinfo {author} {\bibfnamefont {I.~N.}\ \bibnamefont
  {Filikhin}}\ and\ \bibinfo {author} {\bibfnamefont {A.}~\bibnamefont {Gal}},\
  }\href {\doibase 10.1103/PhysRevLett.89.172502} {\bibfield  {journal}
  {\bibinfo  {journal} {Phys. Rev. Lett.}\ }\textbf {\bibinfo {volume} {89}},\
  \bibinfo {pages} {172502} (\bibinfo {year} {2002})},\ \Eprint
  {http://arxiv.org/abs/nucl-th/0209003} {arXiv:nucl-th/0209003} \BibitemShut
  {NoStop}%
\bibitem [{\citenamefont {Garcilazo}\ \emph {et~al.}(2007)\citenamefont
  {Garcilazo}, \citenamefont {Valcarce},\ and\ \citenamefont
  {Fernandez-Carames}}]{Garcilazo:2007pc}%
  \BibitemOpen
  \bibfield  {author} {\bibinfo {author} {\bibfnamefont {H.}~\bibnamefont
  {Garcilazo}}, \bibinfo {author} {\bibfnamefont {A.}~\bibnamefont {Valcarce}},
  \ and\ \bibinfo {author} {\bibfnamefont {T.}~\bibnamefont
  {Fernandez-Carames}},\ }\href {\doibase 10.1103/PhysRevC.76.034001}
  {\bibfield  {journal} {\bibinfo  {journal} {Phys. Rev. C}\ }\textbf {\bibinfo
  {volume} {76}},\ \bibinfo {pages} {034001} (\bibinfo {year} {2007})},\
  \Eprint {http://arxiv.org/abs/0708.0199} {arXiv:0708.0199 [hep-ph]}
  \BibitemShut {NoStop}%
\bibitem [{\citenamefont {Contessi}\ \emph {et~al.}(2018)\citenamefont
  {Contessi}, \citenamefont {Barnea},\ and\ \citenamefont
  {Gal}}]{Contessi:2018qnz}%
  \BibitemOpen
  \bibfield  {author} {\bibinfo {author} {\bibfnamefont {L.}~\bibnamefont
  {Contessi}}, \bibinfo {author} {\bibfnamefont {N.}~\bibnamefont {Barnea}}, \
  and\ \bibinfo {author} {\bibfnamefont {A.}~\bibnamefont {Gal}},\ }\href
  {\doibase 10.1103/PhysRevLett.121.102502} {\bibfield  {journal} {\bibinfo
  {journal} {Phys. Rev. Lett.}\ }\textbf {\bibinfo {volume} {121}},\ \bibinfo
  {pages} {102502} (\bibinfo {year} {2018})},\ \Eprint
  {http://arxiv.org/abs/1805.04302} {arXiv:1805.04302 [nucl-th]} \BibitemShut
  {NoStop}%
\bibitem [{\citenamefont {Lonardoni}\ \emph {et~al.}(2013)\citenamefont
  {Lonardoni}, \citenamefont {Gandolfi},\ and\ \citenamefont
  {Pederiva}}]{Lonardoni:2013rm}%
  \BibitemOpen
  \bibfield  {author} {\bibinfo {author} {\bibfnamefont {D.}~\bibnamefont
  {Lonardoni}}, \bibinfo {author} {\bibfnamefont {S.}~\bibnamefont {Gandolfi}},
  \ and\ \bibinfo {author} {\bibfnamefont {F.}~\bibnamefont {Pederiva}},\
  }\href {\doibase 10.1103/PhysRevC.87.041303} {\bibfield  {journal} {\bibinfo
  {journal} {Phys. Rev. C}\ }\textbf {\bibinfo {volume} {87}},\ \bibinfo
  {pages} {041303} (\bibinfo {year} {2013})},\ \Eprint
  {http://arxiv.org/abs/1301.7472} {arXiv:1301.7472 [nucl-th]} \BibitemShut
  {NoStop}%
\bibitem [{\citenamefont {Lonardoni}\ \emph {et~al.}(2014)\citenamefont
  {Lonardoni}, \citenamefont {Pederiva},\ and\ \citenamefont
  {Gandolfi}}]{Lonardoni:2013gta}%
  \BibitemOpen
  \bibfield  {author} {\bibinfo {author} {\bibfnamefont {D.}~\bibnamefont
  {Lonardoni}}, \bibinfo {author} {\bibfnamefont {F.}~\bibnamefont {Pederiva}},
  \ and\ \bibinfo {author} {\bibfnamefont {S.}~\bibnamefont {Gandolfi}},\
  }\href {\doibase 10.1103/PhysRevC.89.014314} {\bibfield  {journal} {\bibinfo
  {journal} {Phys. Rev. C}\ }\textbf {\bibinfo {volume} {89}},\ \bibinfo
  {pages} {014314} (\bibinfo {year} {2014})},\ \Eprint
  {http://arxiv.org/abs/1312.3844} {arXiv:1312.3844 [nucl-th]} \BibitemShut
  {NoStop}%
\bibitem [{\citenamefont {Wirth}\ \emph {et~al.}(2014)\citenamefont {Wirth},
  \citenamefont {Gazda}, \citenamefont {Navr\'atil}, \citenamefont {Calci},
  \citenamefont {Langhammer},\ and\ \citenamefont {Roth}}]{Wirth:2014apa}%
  \BibitemOpen
  \bibfield  {author} {\bibinfo {author} {\bibfnamefont {R.}~\bibnamefont
  {Wirth}}, \bibinfo {author} {\bibfnamefont {D.}~\bibnamefont {Gazda}},
  \bibinfo {author} {\bibfnamefont {P.}~\bibnamefont {Navr\'atil}}, \bibinfo
  {author} {\bibfnamefont {A.}~\bibnamefont {Calci}}, \bibinfo {author}
  {\bibfnamefont {J.}~\bibnamefont {Langhammer}}, \ and\ \bibinfo {author}
  {\bibfnamefont {R.}~\bibnamefont {Roth}},\ }\href {\doibase
  10.1103/PhysRevLett.113.192502} {\bibfield  {journal} {\bibinfo  {journal}
  {Phys. Rev. Lett.}\ }\textbf {\bibinfo {volume} {113}},\ \bibinfo {pages}
  {192502} (\bibinfo {year} {2014})},\ \Eprint {http://arxiv.org/abs/1403.3067}
  {arXiv:1403.3067 [nucl-th]} \BibitemShut {NoStop}%
\bibitem [{\citenamefont {Wirth}\ and\ \citenamefont
  {Roth}(2016)}]{Wirth:2016iwn}%
  \BibitemOpen
  \bibfield  {author} {\bibinfo {author} {\bibfnamefont {R.}~\bibnamefont
  {Wirth}}\ and\ \bibinfo {author} {\bibfnamefont {R.}~\bibnamefont {Roth}},\
  }\href {\doibase 10.1103/PhysRevLett.117.182501} {\bibfield  {journal}
  {\bibinfo  {journal} {Phys. Rev. Lett.}\ }\textbf {\bibinfo {volume} {117}},\
  \bibinfo {pages} {182501} (\bibinfo {year} {2016})},\ \Eprint
  {http://arxiv.org/abs/1605.08677} {arXiv:1605.08677 [nucl-th]} \BibitemShut
  {NoStop}%
\bibitem [{\citenamefont {Wirth}\ and\ \citenamefont
  {Roth}(2018)}]{Wirth:2017lso}%
  \BibitemOpen
  \bibfield  {author} {\bibinfo {author} {\bibfnamefont {R.}~\bibnamefont
  {Wirth}}\ and\ \bibinfo {author} {\bibfnamefont {R.}~\bibnamefont {Roth}},\
  }\href {\doibase 10.1016/j.physletb.2018.02.021} {\bibfield  {journal}
  {\bibinfo  {journal} {Phys. Lett. B}\ }\textbf {\bibinfo {volume} {779}},\
  \bibinfo {pages} {336} (\bibinfo {year} {2018})},\ \Eprint
  {http://arxiv.org/abs/1710.04880} {arXiv:1710.04880 [nucl-th]} \BibitemShut
  {NoStop}%
\bibitem [{\citenamefont {Wirth}\ and\ \citenamefont
  {Roth}(2019)}]{Wirth:2019cpp}%
  \BibitemOpen
  \bibfield  {author} {\bibinfo {author} {\bibfnamefont {R.}~\bibnamefont
  {Wirth}}\ and\ \bibinfo {author} {\bibfnamefont {R.}~\bibnamefont {Roth}},\
  }\href {\doibase 10.1103/PhysRevC.100.044313} {\bibfield  {journal} {\bibinfo
   {journal} {Phys. Rev. C}\ }\textbf {\bibinfo {volume} {100}},\ \bibinfo
  {pages} {044313} (\bibinfo {year} {2019})},\ \Eprint
  {http://arxiv.org/abs/1902.03324} {arXiv:1902.03324 [nucl-th]} \BibitemShut
  {NoStop}%
\bibitem [{\citenamefont {Kn\"oll}\ and\ \citenamefont
  {Roth}(2023)}]{Knoll:2023mqk}%
  \BibitemOpen
  \bibfield  {author} {\bibinfo {author} {\bibfnamefont {M.}~\bibnamefont
  {Kn\"oll}}\ and\ \bibinfo {author} {\bibfnamefont {R.}~\bibnamefont {Roth}},\
  }\href {\doibase 10.1016/j.physletb.2023.138258} {\bibfield  {journal}
  {\bibinfo  {journal} {Phys. Lett. B}\ }\textbf {\bibinfo {volume} {846}},\
  \bibinfo {pages} {138258} (\bibinfo {year} {2023})},\ \Eprint
  {http://arxiv.org/abs/2307.11577} {arXiv:2307.11577 [nucl-th]} \BibitemShut
  {NoStop}%
\bibitem [{\citenamefont {Le}\ \emph {et~al.}(2025)\citenamefont {Le},
  \citenamefont {Haidenbauer}, \citenamefont {Mei\ss{}ner},\ and\ \citenamefont
  {Nogga}}]{Le:2024rkd}%
  \BibitemOpen
  \bibfield  {author} {\bibinfo {author} {\bibfnamefont {H.}~\bibnamefont
  {Le}}, \bibinfo {author} {\bibfnamefont {J.}~\bibnamefont {Haidenbauer}},
  \bibinfo {author} {\bibfnamefont {U.-G.}\ \bibnamefont {Mei\ss{}ner}}, \ and\
  \bibinfo {author} {\bibfnamefont {A.}~\bibnamefont {Nogga}},\ }\href
  {\doibase 10.1103/PhysRevLett.134.072502} {\bibfield  {journal} {\bibinfo
  {journal} {Phys. Rev. Lett.}\ }\textbf {\bibinfo {volume} {134}},\ \bibinfo
  {pages} {072502} (\bibinfo {year} {2025})},\ \Eprint
  {http://arxiv.org/abs/2409.18577} {arXiv:2409.18577 [nucl-th]} \BibitemShut
  {NoStop}%
\bibitem [{\citenamefont {Adams}\ \emph {et~al.}(2021)\citenamefont {Adams},
  \citenamefont {Carleo}, \citenamefont {Lovato},\ and\ \citenamefont
  {Rocco}}]{Adams:2020aax}%
  \BibitemOpen
  \bibfield  {author} {\bibinfo {author} {\bibfnamefont {C.}~\bibnamefont
  {Adams}}, \bibinfo {author} {\bibfnamefont {G.}~\bibnamefont {Carleo}},
  \bibinfo {author} {\bibfnamefont {A.}~\bibnamefont {Lovato}}, \ and\ \bibinfo
  {author} {\bibfnamefont {N.}~\bibnamefont {Rocco}},\ }\href {\doibase
  10.1103/PhysRevLett.127.022502} {\bibfield  {journal} {\bibinfo  {journal}
  {Phys. Rev. Lett.}\ }\textbf {\bibinfo {volume} {127}},\ \bibinfo {pages}
  {022502} (\bibinfo {year} {2021})},\ \Eprint
  {http://arxiv.org/abs/2007.14282} {arXiv:2007.14282 [nucl-th]} \BibitemShut
  {NoStop}%
\bibitem [{\citenamefont {Lovato}\ \emph {et~al.}(2022)\citenamefont {Lovato},
  \citenamefont {Adams}, \citenamefont {Carleo},\ and\ \citenamefont
  {Rocco}}]{Lovato:2022tjh}%
  \BibitemOpen
  \bibfield  {author} {\bibinfo {author} {\bibfnamefont {A.}~\bibnamefont
  {Lovato}}, \bibinfo {author} {\bibfnamefont {C.}~\bibnamefont {Adams}},
  \bibinfo {author} {\bibfnamefont {G.}~\bibnamefont {Carleo}}, \ and\ \bibinfo
  {author} {\bibfnamefont {N.}~\bibnamefont {Rocco}},\ }\href {\doibase
  10.1103/PhysRevResearch.4.043178} {\bibfield  {journal} {\bibinfo  {journal}
  {Phys. Rev. Res.}\ }\textbf {\bibinfo {volume} {4}},\ \bibinfo {pages}
  {043178} (\bibinfo {year} {2022})},\ \Eprint
  {http://arxiv.org/abs/2206.10021} {arXiv:2206.10021 [nucl-th]} \BibitemShut
  {NoStop}%
\bibitem [{\citenamefont {Kim}\ \emph {et~al.}(2024)\citenamefont {Kim},
  \citenamefont {Pescia}, \citenamefont {Fore}, \citenamefont {Nys},
  \citenamefont {Carleo}, \citenamefont {Gandolfi}, \citenamefont
  {Hjorth-Jensen},\ and\ \citenamefont {Lovato}}]{Kim:2023fwy}%
  \BibitemOpen
  \bibfield  {author} {\bibinfo {author} {\bibfnamefont {J.}~\bibnamefont
  {Kim}}, \bibinfo {author} {\bibfnamefont {G.}~\bibnamefont {Pescia}},
  \bibinfo {author} {\bibfnamefont {B.}~\bibnamefont {Fore}}, \bibinfo {author}
  {\bibfnamefont {J.}~\bibnamefont {Nys}}, \bibinfo {author} {\bibfnamefont
  {G.}~\bibnamefont {Carleo}}, \bibinfo {author} {\bibfnamefont
  {S.}~\bibnamefont {Gandolfi}}, \bibinfo {author} {\bibfnamefont
  {M.}~\bibnamefont {Hjorth-Jensen}}, \ and\ \bibinfo {author} {\bibfnamefont
  {A.}~\bibnamefont {Lovato}},\ }\href {\doibase 10.1038/s42005-024-01613-w}
  {\bibfield  {journal} {\bibinfo  {journal} {Commun. Phys.}\ }\textbf
  {\bibinfo {volume} {7}},\ \bibinfo {pages} {148} (\bibinfo {year} {2024})},\
  \Eprint {http://arxiv.org/abs/2305.08831} {arXiv:2305.08831
  [cond-mat.quant-gas]} \BibitemShut {NoStop}%
\bibitem [{\citenamefont {Fore}\ \emph {et~al.}(2025)\citenamefont {Fore},
  \citenamefont {Kim}, \citenamefont {Hjorth-Jensen},\ and\ \citenamefont
  {Lovato}}]{Fore:2024exa}%
  \BibitemOpen
  \bibfield  {author} {\bibinfo {author} {\bibfnamefont {B.}~\bibnamefont
  {Fore}}, \bibinfo {author} {\bibfnamefont {J.}~\bibnamefont {Kim}}, \bibinfo
  {author} {\bibfnamefont {M.}~\bibnamefont {Hjorth-Jensen}}, \ and\ \bibinfo
  {author} {\bibfnamefont {A.}~\bibnamefont {Lovato}},\ }\href {\doibase
  10.1038/s42005-025-02015-2} {\bibfield  {journal} {\bibinfo  {journal}
  {Commun. Phys.}\ }\textbf {\bibinfo {volume} {8}},\ \bibinfo {pages} {108}
  (\bibinfo {year} {2025})},\ \Eprint {http://arxiv.org/abs/2407.21207}
  {arXiv:2407.21207 [nucl-th]} \BibitemShut {NoStop}%
\bibitem [{\citenamefont {Schiavilla}\ \emph {et~al.}(2021)\citenamefont
  {Schiavilla}, \citenamefont {Girlanda}, \citenamefont {Gnech}, \citenamefont
  {Kievsky}, \citenamefont {Lovato}, \citenamefont {Marcucci}, \citenamefont
  {Piarulli},\ and\ \citenamefont {Viviani}}]{Schiavilla:2021dun}%
  \BibitemOpen
  \bibfield  {author} {\bibinfo {author} {\bibfnamefont {R.}~\bibnamefont
  {Schiavilla}}, \bibinfo {author} {\bibfnamefont {L.}~\bibnamefont
  {Girlanda}}, \bibinfo {author} {\bibfnamefont {A.}~\bibnamefont {Gnech}},
  \bibinfo {author} {\bibfnamefont {A.}~\bibnamefont {Kievsky}}, \bibinfo
  {author} {\bibfnamefont {A.}~\bibnamefont {Lovato}}, \bibinfo {author}
  {\bibfnamefont {L.~E.}\ \bibnamefont {Marcucci}}, \bibinfo {author}
  {\bibfnamefont {M.}~\bibnamefont {Piarulli}}, \ and\ \bibinfo {author}
  {\bibfnamefont {M.}~\bibnamefont {Viviani}},\ }\href {\doibase
  10.1103/PhysRevC.103.054003} {\bibfield  {journal} {\bibinfo  {journal}
  {Phys. Rev. C}\ }\textbf {\bibinfo {volume} {103}},\ \bibinfo {pages}
  {054003} (\bibinfo {year} {2021})},\ \Eprint
  {http://arxiv.org/abs/2102.02327} {arXiv:2102.02327 [nucl-th]} \BibitemShut
  {NoStop}%
\bibitem [{\citenamefont {Auerbach}\ \emph {et~al.}(1972)\citenamefont
  {Auerbach}, \citenamefont {Hüfner}, \citenamefont {Kerman},\ and\
  \citenamefont {Shakin}}]{auerbachTheoryIsobaricAnalog1972}%
  \BibitemOpen
  \bibfield  {author} {\bibinfo {author} {\bibfnamefont {N.}~\bibnamefont
  {Auerbach}}, \bibinfo {author} {\bibfnamefont {J.}~\bibnamefont {Hüfner}},
  \bibinfo {author} {\bibfnamefont {A.~K.}\ \bibnamefont {Kerman}}, \ and\
  \bibinfo {author} {\bibfnamefont {C.~M.}\ \bibnamefont {Shakin}},\ }\href
  {\doibase 10.1103/RevModPhys.44.48} {\bibfield  {journal} {\bibinfo
  {journal} {Reviews of Modern Physics}\ }\textbf {\bibinfo {volume} {44}},\
  \bibinfo {pages} {48} (\bibinfo {year} {1972})}\BibitemShut {NoStop}%
\bibitem [{\citenamefont {Hackenburg}(2006)}]{Hackenburg:2006qd}%
  \BibitemOpen
  \bibfield  {author} {\bibinfo {author} {\bibfnamefont {R.~W.}\ \bibnamefont
  {Hackenburg}},\ }\href {\doibase 10.1103/PhysRevC.73.044002} {\bibfield
  {journal} {\bibinfo  {journal} {Phys. Rev. C}\ }\textbf {\bibinfo {volume}
  {73}},\ \bibinfo {pages} {044002} (\bibinfo {year} {2006})}\BibitemShut
  {NoStop}%
\bibitem [{\citenamefont {Alexander}\ \emph {et~al.}(1968)\citenamefont
  {Alexander}, \citenamefont {Karshon}, \citenamefont {Shapira}, \citenamefont
  {Yekutieli}, \citenamefont {Engelmann}, \citenamefont {Filthuth},\ and\
  \citenamefont {Lughofer}}]{Alexander:1968acu}%
  \BibitemOpen
  \bibfield  {author} {\bibinfo {author} {\bibfnamefont {G.}~\bibnamefont
  {Alexander}}, \bibinfo {author} {\bibfnamefont {U.}~\bibnamefont {Karshon}},
  \bibinfo {author} {\bibfnamefont {A.}~\bibnamefont {Shapira}}, \bibinfo
  {author} {\bibfnamefont {G.}~\bibnamefont {Yekutieli}}, \bibinfo {author}
  {\bibfnamefont {R.}~\bibnamefont {Engelmann}}, \bibinfo {author}
  {\bibfnamefont {H.}~\bibnamefont {Filthuth}}, \ and\ \bibinfo {author}
  {\bibfnamefont {W.}~\bibnamefont {Lughofer}},\ }\href {\doibase
  10.1103/PhysRev.173.1452} {\bibfield  {journal} {\bibinfo  {journal} {Phys.
  Rev.}\ }\textbf {\bibinfo {volume} {173}},\ \bibinfo {pages} {1452} (\bibinfo
  {year} {1968})}\BibitemShut {NoStop}%
\bibitem [{\citenamefont {Bedaque}\ and\ \citenamefont {van
  Kolck}(2002)}]{Bedaque:2002mn}%
  \BibitemOpen
  \bibfield  {author} {\bibinfo {author} {\bibfnamefont {P.~F.}\ \bibnamefont
  {Bedaque}}\ and\ \bibinfo {author} {\bibfnamefont {U.}~\bibnamefont {van
  Kolck}},\ }\href {\doibase 10.1146/annurev.nucl.52.050102.090637} {\bibfield
  {journal} {\bibinfo  {journal} {Ann. Rev. Nucl. Part. Sci.}\ }\textbf
  {\bibinfo {volume} {52}},\ \bibinfo {pages} {339} (\bibinfo {year} {2002})},\
  \Eprint {http://arxiv.org/abs/nucl-th/0203055} {arXiv:nucl-th/0203055}
  \BibitemShut {NoStop}%
\bibitem [{\citenamefont {Stetcu}\ \emph {et~al.}(2007)\citenamefont {Stetcu},
  \citenamefont {Barrett},\ and\ \citenamefont {van Kolck}}]{Stetcu:2006ey}%
  \BibitemOpen
  \bibfield  {author} {\bibinfo {author} {\bibfnamefont {I.}~\bibnamefont
  {Stetcu}}, \bibinfo {author} {\bibfnamefont {B.~R.}\ \bibnamefont {Barrett}},
  \ and\ \bibinfo {author} {\bibfnamefont {U.}~\bibnamefont {van Kolck}},\
  }\href {\doibase 10.1016/j.physletb.2007.07.065} {\bibfield  {journal}
  {\bibinfo  {journal} {Phys. Lett. B}\ }\textbf {\bibinfo {volume} {653}},\
  \bibinfo {pages} {358} (\bibinfo {year} {2007})},\ \Eprint
  {http://arxiv.org/abs/nucl-th/0609023} {arXiv:nucl-th/0609023} \BibitemShut
  {NoStop}%
\bibitem [{\citenamefont {Contessi}\ \emph {et~al.}(2017)\citenamefont
  {Contessi}, \citenamefont {Lovato}, \citenamefont {Pederiva}, \citenamefont
  {Roggero}, \citenamefont {Kirscher},\ and\ \citenamefont {van
  Kolck}}]{Contessi:2017rww}%
  \BibitemOpen
  \bibfield  {author} {\bibinfo {author} {\bibfnamefont {L.}~\bibnamefont
  {Contessi}}, \bibinfo {author} {\bibfnamefont {A.}~\bibnamefont {Lovato}},
  \bibinfo {author} {\bibfnamefont {F.}~\bibnamefont {Pederiva}}, \bibinfo
  {author} {\bibfnamefont {A.}~\bibnamefont {Roggero}}, \bibinfo {author}
  {\bibfnamefont {J.}~\bibnamefont {Kirscher}}, \ and\ \bibinfo {author}
  {\bibfnamefont {U.}~\bibnamefont {van Kolck}},\ }\href {\doibase
  10.1016/j.physletb.2017.07.048} {\bibfield  {journal} {\bibinfo  {journal}
  {Phys. Lett. B}\ }\textbf {\bibinfo {volume} {772}},\ \bibinfo {pages} {839}
  (\bibinfo {year} {2017})},\ \Eprint {http://arxiv.org/abs/1701.06516}
  {arXiv:1701.06516 [nucl-th]} \BibitemShut {NoStop}%
\bibitem [{\citenamefont {Bansal}\ \emph {et~al.}(2018)\citenamefont {Bansal},
  \citenamefont {Binder}, \citenamefont {Ekstr\"om}, \citenamefont {Hagen},
  \citenamefont {Jansen},\ and\ \citenamefont {Papenbrock}}]{Bansal:2017pwn}%
  \BibitemOpen
  \bibfield  {author} {\bibinfo {author} {\bibfnamefont {A.}~\bibnamefont
  {Bansal}}, \bibinfo {author} {\bibfnamefont {S.}~\bibnamefont {Binder}},
  \bibinfo {author} {\bibfnamefont {A.}~\bibnamefont {Ekstr\"om}}, \bibinfo
  {author} {\bibfnamefont {G.}~\bibnamefont {Hagen}}, \bibinfo {author}
  {\bibfnamefont {G.~R.}\ \bibnamefont {Jansen}}, \ and\ \bibinfo {author}
  {\bibfnamefont {T.}~\bibnamefont {Papenbrock}},\ }\href {\doibase
  10.1103/PhysRevC.98.054301} {\bibfield  {journal} {\bibinfo  {journal} {Phys.
  Rev. C}\ }\textbf {\bibinfo {volume} {98}},\ \bibinfo {pages} {054301}
  (\bibinfo {year} {2018})},\ \Eprint {http://arxiv.org/abs/1712.10246}
  {arXiv:1712.10246 [nucl-th]} \BibitemShut {NoStop}%
\bibitem [{\citenamefont {Lu}\ \emph {et~al.}(2019)\citenamefont {Lu},
  \citenamefont {Li}, \citenamefont {Elhatisari}, \citenamefont {Lee},
  \citenamefont {Epelbaum},\ and\ \citenamefont {Mei\ss{}ner}}]{Lu:2018bat}%
  \BibitemOpen
  \bibfield  {author} {\bibinfo {author} {\bibfnamefont {B.-N.}\ \bibnamefont
  {Lu}}, \bibinfo {author} {\bibfnamefont {N.}~\bibnamefont {Li}}, \bibinfo
  {author} {\bibfnamefont {S.}~\bibnamefont {Elhatisari}}, \bibinfo {author}
  {\bibfnamefont {D.}~\bibnamefont {Lee}}, \bibinfo {author} {\bibfnamefont
  {E.}~\bibnamefont {Epelbaum}}, \ and\ \bibinfo {author} {\bibfnamefont
  {U.-G.}\ \bibnamefont {Mei\ss{}ner}},\ }\href {\doibase
  10.1016/j.physletb.2019.134863} {\bibfield  {journal} {\bibinfo  {journal}
  {Phys. Lett. B}\ }\textbf {\bibinfo {volume} {797}},\ \bibinfo {pages}
  {134863} (\bibinfo {year} {2019})},\ \Eprint
  {http://arxiv.org/abs/1812.10928} {arXiv:1812.10928 [nucl-th]} \BibitemShut
  {NoStop}%
\bibitem [{\citenamefont {Gnech}\ \emph {et~al.}(2024)\citenamefont {Gnech},
  \citenamefont {Fore}, \citenamefont {Tropiano},\ and\ \citenamefont
  {Lovato}}]{Gnech:2023prs}%
  \BibitemOpen
  \bibfield  {author} {\bibinfo {author} {\bibfnamefont {A.}~\bibnamefont
  {Gnech}}, \bibinfo {author} {\bibfnamefont {B.}~\bibnamefont {Fore}},
  \bibinfo {author} {\bibfnamefont {A.~J.}\ \bibnamefont {Tropiano}}, \ and\
  \bibinfo {author} {\bibfnamefont {A.}~\bibnamefont {Lovato}},\ }\href
  {\doibase 10.1103/PhysRevLett.133.142501} {\bibfield  {journal} {\bibinfo
  {journal} {Phys. Rev. Lett.}\ }\textbf {\bibinfo {volume} {133}},\ \bibinfo
  {pages} {142501} (\bibinfo {year} {2024})},\ \Eprint
  {http://arxiv.org/abs/2308.16266} {arXiv:2308.16266 [nucl-th]} \BibitemShut
  {NoStop}%
\bibitem [{\citenamefont {Contessi}\ \emph
  {et~al.}(2024{\natexlab{a}})\citenamefont {Contessi}, \citenamefont
  {Sch\"afer},\ and\ \citenamefont {van Kolck}}]{Contessi:2023yoz}%
  \BibitemOpen
  \bibfield  {author} {\bibinfo {author} {\bibfnamefont {L.}~\bibnamefont
  {Contessi}}, \bibinfo {author} {\bibfnamefont {M.}~\bibnamefont {Sch\"afer}},
  \ and\ \bibinfo {author} {\bibfnamefont {U.}~\bibnamefont {van Kolck}},\
  }\href {\doibase 10.1103/PhysRevA.109.022814} {\bibfield  {journal} {\bibinfo
   {journal} {Phys. Rev. A}\ }\textbf {\bibinfo {volume} {109}},\ \bibinfo
  {pages} {022814} (\bibinfo {year} {2024}{\natexlab{a}})},\ \Eprint
  {http://arxiv.org/abs/2310.15760} {arXiv:2310.15760 [physics.atm-clus]}
  \BibitemShut {NoStop}%
\bibitem [{\citenamefont {Contessi}\ \emph
  {et~al.}(2024{\natexlab{b}})\citenamefont {Contessi}, \citenamefont
  {Pavon~Valderrama},\ and\ \citenamefont {van Kolck}}]{Contessi:2024vae}%
  \BibitemOpen
  \bibfield  {author} {\bibinfo {author} {\bibfnamefont {L.}~\bibnamefont
  {Contessi}}, \bibinfo {author} {\bibfnamefont {M.}~\bibnamefont
  {Pavon~Valderrama}}, \ and\ \bibinfo {author} {\bibfnamefont
  {U.}~\bibnamefont {van Kolck}},\ }\href {\doibase
  10.1016/j.physletb.2024.138903} {\bibfield  {journal} {\bibinfo  {journal}
  {Phys. Lett. B}\ }\textbf {\bibinfo {volume} {856}},\ \bibinfo {pages}
  {138903} (\bibinfo {year} {2024}{\natexlab{b}})},\ \Eprint
  {http://arxiv.org/abs/2403.16596} {arXiv:2403.16596 [nucl-th]} \BibitemShut
  {NoStop}%
\bibitem [{\citenamefont {Contessi}\ \emph {et~al.}(2025)\citenamefont
  {Contessi}, \citenamefont {Sch\"afer}, \citenamefont {Gnech}, \citenamefont
  {Lovato},\ and\ \citenamefont {van Kolck}}]{Contessi:2025xue}%
  \BibitemOpen
  \bibfield  {author} {\bibinfo {author} {\bibfnamefont {L.}~\bibnamefont
  {Contessi}}, \bibinfo {author} {\bibfnamefont {M.}~\bibnamefont {Sch\"afer}},
  \bibinfo {author} {\bibfnamefont {A.}~\bibnamefont {Gnech}}, \bibinfo
  {author} {\bibfnamefont {A.}~\bibnamefont {Lovato}}, \ and\ \bibinfo {author}
  {\bibfnamefont {U.}~\bibnamefont {van Kolck}},\ }\href@noop {} {\  (\bibinfo
  {year} {2025})},\ \Eprint {http://arxiv.org/abs/2505.09299} {arXiv:2505.09299
  [nucl-th]} \BibitemShut {NoStop}%
\bibitem [{\citenamefont
  {Calogero}(1963)}]{calogeroNovelApproachElementary1963}%
  \BibitemOpen
  \bibfield  {author} {\bibinfo {author} {\bibfnamefont {F.}~\bibnamefont
  {Calogero}},\ }\href {\doibase 10.1007/BF02812620} {\bibfield  {journal}
  {\bibinfo  {journal} {Il Nuovo Cimento}\ }\textbf {\bibinfo {volume} {27}},\
  \bibinfo {pages} {261} (\bibinfo {year} {1963})}\BibitemShut {NoStop}%
\bibitem [{\citenamefont {{International Atomic Energy Agency
  (IAEA)}}()}]{iaea_chart}%
  \BibitemOpen
  \bibfield  {author} {\bibinfo {author} {\bibnamefont {{International Atomic
  Energy Agency (IAEA)}}},\ }\href@noop {} {\enquote {\bibinfo {title} {Live
  chart of nuclides},}\ }\bibinfo {howpublished}
  {\url{https://www-nds.iaea.org/relnsd/vcharthtml/VChartHTML.html}}\BibitemShut
  {NoStop}%
\bibitem [{\citenamefont {Ekstr\"om}\ and\ \citenamefont
  {Platter}(2025)}]{Ekstrom:2024dqr}%
  \BibitemOpen
  \bibfield  {author} {\bibinfo {author} {\bibfnamefont {A.}~\bibnamefont
  {Ekstr\"om}}\ and\ \bibinfo {author} {\bibfnamefont {L.}~\bibnamefont
  {Platter}},\ }\href {\doibase 10.1016/j.physletb.2024.139207} {\bibfield
  {journal} {\bibinfo  {journal} {Phys. Lett. B}\ }\textbf {\bibinfo {volume}
  {860}},\ \bibinfo {pages} {139207} (\bibinfo {year} {2025})},\ \Eprint
  {http://arxiv.org/abs/2409.08197} {arXiv:2409.08197 [nucl-th]} \BibitemShut
  {NoStop}%
\bibitem [{\citenamefont {Eckert}\ and\ \citenamefont
  {Achenbach}(2023)}]{eckertChartHypernuclidesHypernuclear2023}%
  \BibitemOpen
  \bibfield  {author} {\bibinfo {author} {\bibfnamefont {P.}~\bibnamefont
  {Eckert}}\ and\ \bibinfo {author} {\bibfnamefont {P.}~\bibnamefont
  {Achenbach}},\ }\href@noop {} {\enquote {\bibinfo {title} {Chart of
  {{Hypernuclides}} --- {{Hypernuclear Structure}} and {{Decay Data}}},}\
  }\bibinfo {howpublished} {https://hypernuclei.kph.uni-mainz.de/} (\bibinfo
  {year} {2023})\BibitemShut {NoStop}%
\bibitem [{\citenamefont {Rasmussen}\ and\ \citenamefont
  {Williams}(2008)}]{rasmussenGaussianProcessesMachine2008}%
  \BibitemOpen
  \bibfield  {author} {\bibinfo {author} {\bibfnamefont {C.~E.}\ \bibnamefont
  {Rasmussen}}\ and\ \bibinfo {author} {\bibfnamefont {C.~K.~I.}\ \bibnamefont
  {Williams}},\ }\href@noop {} {\emph {\bibinfo {title} {Gaussian Processes for
  Machine Learning}}},\ \bibinfo {edition} {3rd}\ ed.,\ Adaptive Computation
  and Machine Learning\ (\bibinfo  {publisher} {MIT Press},\ \bibinfo {address}
  {Cambridge, Mass.},\ \bibinfo {year} {2008})\BibitemShut {NoStop}%
\bibitem [{\citenamefont {Suzuki}\ and\ \citenamefont
  {Varga}(1998)}]{suzukiStochasticVariationalApproach1998c}%
  \BibitemOpen
  \bibfield  {author} {\bibinfo {author} {\bibfnamefont {Y.}~\bibnamefont
  {Suzuki}}\ and\ \bibinfo {author} {\bibfnamefont {K.}~\bibnamefont {Varga}},\
  }\href@noop {} {\emph {\bibinfo {title} {Stochastic Variational Approach to
  Quantum Mechanical Few-Body Problems}}},\ \bibinfo {series} {Lecture Notes in
  Physics {{New}} Series {{M}}, Monographs}\ No.~\bibinfo {number} {54}\
  (\bibinfo  {publisher} {Springer},\ \bibinfo {address} {Berlin Heidelberg New
  York Barcelona Hong Kong London Milan Paris Singapore Tokyo},\ \bibinfo
  {year} {1998})\BibitemShut {NoStop}%
\bibitem [{\citenamefont {Duvenaud}(2014)}]{Duvenaud2014}%
  \BibitemOpen
  \bibfield  {author} {\bibinfo {author} {\bibfnamefont {D.}~\bibnamefont
  {Duvenaud}},\ }\emph {\bibinfo {title} {Automatic model construction with
  Gaussian processes}},\ \href {\doibase 10.17863/CAM.14087} {Ph.D. thesis},\
  \bibinfo  {school} {University of Cambridge} (\bibinfo {year}
  {2014})\BibitemShut {NoStop}%
\bibitem [{\citenamefont
  {MacKay}(1996)}]{mackayBayesianNonLinearModeling1996a}%
  \BibitemOpen
  \bibfield  {author} {\bibinfo {author} {\bibfnamefont {D.~J.~C.}\
  \bibnamefont {MacKay}},\ }in\ \href {\doibase 10.1007/978-94-015-8729-7_18}
  {\emph {\bibinfo {booktitle} {Maximum {{Entropy}} and {{Bayesian
  Methods}}}}},\ \bibinfo {editor} {edited by\ \bibinfo {editor} {\bibfnamefont
  {G.~R.}\ \bibnamefont {Heidbreder}}}\ (\bibinfo  {publisher} {Springer
  Netherlands},\ \bibinfo {address} {Dordrecht},\ \bibinfo {year} {1996})\ pp.\
  \bibinfo {pages} {221--234}\BibitemShut {NoStop}%
\bibitem [{\citenamefont {{ALICE
  Collaboration}}(2023)}]{alicecollaborationMeasurementLifetimeLambda2023a}%
  \BibitemOpen
  \bibfield  {author} {\bibinfo {author} {\bibnamefont {{ALICE
  Collaboration}}},\ }\href {\doibase 10.1103/PhysRevLett.131.102302}
  {\bibfield  {journal} {\bibinfo  {journal} {Physical Review Letters}\
  }\textbf {\bibinfo {volume} {131}},\ \bibinfo {pages} {102302} (\bibinfo
  {year} {2023})},\ \Eprint {http://arxiv.org/abs/2209.07360} {arXiv:2209.07360
  [nucl-ex]} \BibitemShut {NoStop}%
\bibitem [{\citenamefont {SheffieldML}(2024)}]{GPyGitHub}%
  \BibitemOpen
  \bibfield  {author} {\bibinfo {author} {\bibnamefont {SheffieldML}},\
  }\href@noop {} {\enquote {\bibinfo {title} {Gpy: A gaussian processes
  framework},}\ }\bibinfo {howpublished}
  {\url{https://github.com/SheffieldML/GPy}} (\bibinfo {year} {2024}),\
  \bibinfo {note} {accessed: 2024-10-10}\BibitemShut {NoStop}%
\bibitem [{\citenamefont {{Foreman-Mackey}}\ \emph {et~al.}(2013)\citenamefont
  {{Foreman-Mackey}}, \citenamefont {Hogg}, \citenamefont {Lang},\ and\
  \citenamefont {Goodman}}]{foreman-mackeyEmceeMCMCHammer2013}%
  \BibitemOpen
  \bibfield  {author} {\bibinfo {author} {\bibfnamefont {D.}~\bibnamefont
  {{Foreman-Mackey}}}, \bibinfo {author} {\bibfnamefont {D.~W.}\ \bibnamefont
  {Hogg}}, \bibinfo {author} {\bibfnamefont {D.}~\bibnamefont {Lang}}, \ and\
  \bibinfo {author} {\bibfnamefont {J.}~\bibnamefont {Goodman}},\ }\href
  {\doibase 10.1086/670067} {\bibfield  {journal} {\bibinfo  {journal}
  {Publications of the Astronomical Society of the Pacific}\ }\textbf {\bibinfo
  {volume} {125}},\ \bibinfo {pages} {306} (\bibinfo {year} {2013})},\ \Eprint
  {http://arxiv.org/abs/1202.3665} {arXiv:1202.3665 [astro-ph]} \BibitemShut
  {NoStop}%
\bibitem [{\citenamefont {Ekstr\"om}\ \emph {et~al.}(2015)\citenamefont
  {Ekstr\"om}, \citenamefont {Jansen}, \citenamefont {Wendt}, \citenamefont
  {Hagen}, \citenamefont {Papenbrock}, \citenamefont {Carlsson}, \citenamefont
  {Forss\'en}, \citenamefont {Hjorth-Jensen}, \citenamefont {Navr\'atil},\ and\
  \citenamefont {Nazarewicz}}]{Ekstrom:2015rta}%
  \BibitemOpen
  \bibfield  {author} {\bibinfo {author} {\bibfnamefont {A.}~\bibnamefont
  {Ekstr\"om}}, \bibinfo {author} {\bibfnamefont {G.~R.}\ \bibnamefont
  {Jansen}}, \bibinfo {author} {\bibfnamefont {K.~A.}\ \bibnamefont {Wendt}},
  \bibinfo {author} {\bibfnamefont {G.}~\bibnamefont {Hagen}}, \bibinfo
  {author} {\bibfnamefont {T.}~\bibnamefont {Papenbrock}}, \bibinfo {author}
  {\bibfnamefont {B.~D.}\ \bibnamefont {Carlsson}}, \bibinfo {author}
  {\bibfnamefont {C.}~\bibnamefont {Forss\'en}}, \bibinfo {author}
  {\bibfnamefont {M.}~\bibnamefont {Hjorth-Jensen}}, \bibinfo {author}
  {\bibfnamefont {P.}~\bibnamefont {Navr\'atil}}, \ and\ \bibinfo {author}
  {\bibfnamefont {W.}~\bibnamefont {Nazarewicz}},\ }\href {\doibase
  10.1103/PhysRevC.109.059901} {\bibfield  {journal} {\bibinfo  {journal}
  {Phys. Rev. C}\ }\textbf {\bibinfo {volume} {91}},\ \bibinfo {pages} {051301}
  (\bibinfo {year} {2015})},\ \bibinfo {note} {[Erratum: Phys.Rev.C 109, 059901
  (2024)]},\ \Eprint {http://arxiv.org/abs/1502.04682} {arXiv:1502.04682
  [nucl-th]} \BibitemShut {NoStop}%
\bibitem [{\citenamefont {{Bukov}}\ \emph {et~al.}(2021)\citenamefont
  {{Bukov}}, \citenamefont {{Schmitt}},\ and\ \citenamefont
  {{Dupont}}}]{bukov:2021}%
  \BibitemOpen
  \bibfield  {author} {\bibinfo {author} {\bibfnamefont {M.}~\bibnamefont
  {{Bukov}}}, \bibinfo {author} {\bibfnamefont {M.}~\bibnamefont {{Schmitt}}},
  \ and\ \bibinfo {author} {\bibfnamefont {M.}~\bibnamefont {{Dupont}}},\
  }\href {\doibase 10.21468/SciPostPhys.10.6.147} {\bibfield  {journal}
  {\bibinfo  {journal} {SciPost Physics}\ }\textbf {\bibinfo {volume} {10}},\
  \bibinfo {eid} {147} (\bibinfo {year} {2021})},\ \Eprint
  {http://arxiv.org/abs/2011.11214} {arXiv:2011.11214 [physics.comp-ph]}
  \BibitemShut {NoStop}%
\bibitem [{\citenamefont {{Zaheer}}\ \emph {et~al.}(2017)\citenamefont
  {{Zaheer}}, \citenamefont {{Kottur}}, \citenamefont {{Ravanbakhsh}},
  \citenamefont {{Poczos}}, \citenamefont {{Salakhutdinov}},\ and\
  \citenamefont {{Smola}}}]{Zaheer:2017}%
  \BibitemOpen
  \bibfield  {author} {\bibinfo {author} {\bibfnamefont {M.}~\bibnamefont
  {{Zaheer}}}, \bibinfo {author} {\bibfnamefont {S.}~\bibnamefont {{Kottur}}},
  \bibinfo {author} {\bibfnamefont {S.}~\bibnamefont {{Ravanbakhsh}}}, \bibinfo
  {author} {\bibfnamefont {B.}~\bibnamefont {{Poczos}}}, \bibinfo {author}
  {\bibfnamefont {R.}~\bibnamefont {{Salakhutdinov}}}, \ and\ \bibinfo {author}
  {\bibfnamefont {A.}~\bibnamefont {{Smola}}},\ }\href@noop {} {\bibfield
  {journal} {\bibinfo  {journal} {arXiv e-prints}\ ,\ \bibinfo {eid}
  {arXiv:1703.06114}} (\bibinfo {year} {2017})},\ \Eprint
  {http://arxiv.org/abs/1703.06114} {arXiv:1703.06114 [cs.LG]} \BibitemShut
  {NoStop}%
\bibitem [{\citenamefont {{Wagstaff}}\ \emph {et~al.}(2019)\citenamefont
  {{Wagstaff}}, \citenamefont {{Fuchs}}, \citenamefont {{Engelcke}},
  \citenamefont {{Posner}},\ and\ \citenamefont {{Osborne}}}]{Wagstaff:2019}%
  \BibitemOpen
  \bibfield  {author} {\bibinfo {author} {\bibfnamefont {E.}~\bibnamefont
  {{Wagstaff}}}, \bibinfo {author} {\bibfnamefont {F.~B.}\ \bibnamefont
  {{Fuchs}}}, \bibinfo {author} {\bibfnamefont {M.}~\bibnamefont {{Engelcke}}},
  \bibinfo {author} {\bibfnamefont {I.}~\bibnamefont {{Posner}}}, \ and\
  \bibinfo {author} {\bibfnamefont {M.}~\bibnamefont {{Osborne}}},\ }\href@noop
  {} {\bibfield  {journal} {\bibinfo  {journal} {arXiv e-prints}\ ,\ \bibinfo
  {eid} {arXiv:1901.09006}} (\bibinfo {year} {2019})},\ \Eprint
  {http://arxiv.org/abs/1901.09006} {arXiv:1901.09006 [cs.LG]} \BibitemShut
  {NoStop}%
\bibitem [{\citenamefont {Feynman}\ and\ \citenamefont
  {Cohen}(1956)}]{Feynman:1956zz}%
  \BibitemOpen
  \bibfield  {author} {\bibinfo {author} {\bibfnamefont {R.~P.}\ \bibnamefont
  {Feynman}}\ and\ \bibinfo {author} {\bibfnamefont {M.}~\bibnamefont
  {Cohen}},\ }\href {\doibase 10.1103/PhysRev.102.1189} {\bibfield  {journal}
  {\bibinfo  {journal} {Phys. Rev.}\ }\textbf {\bibinfo {volume} {102}},\
  \bibinfo {pages} {1189} (\bibinfo {year} {1956})}\BibitemShut {NoStop}%
\bibitem [{\citenamefont {Luo}\ and\ \citenamefont
  {Clark}(2019)}]{Luo:2019iaq}%
  \BibitemOpen
  \bibfield  {author} {\bibinfo {author} {\bibfnamefont {D.}~\bibnamefont
  {Luo}}\ and\ \bibinfo {author} {\bibfnamefont {B.~K.}\ \bibnamefont
  {Clark}},\ }\href {\doibase 10.1103/PhysRevLett.122.226401} {\bibfield
  {journal} {\bibinfo  {journal} {Phys. Rev. Lett.}\ }\textbf {\bibinfo
  {volume} {122}},\ \bibinfo {pages} {226401} (\bibinfo {year}
  {2019})}\BibitemShut {NoStop}%
\bibitem [{\citenamefont {Hermann}\ \emph {et~al.}(2020)\citenamefont
  {Hermann}, \citenamefont {Sch\"atzle},\ and\ \citenamefont
  {No\'e}}]{Hermann:2020xqs}%
  \BibitemOpen
  \bibfield  {author} {\bibinfo {author} {\bibfnamefont {J.}~\bibnamefont
  {Hermann}}, \bibinfo {author} {\bibfnamefont {Z.}~\bibnamefont {Sch\"atzle}},
  \ and\ \bibinfo {author} {\bibfnamefont {F.}~\bibnamefont {No\'e}},\ }\href
  {\doibase 10.1038/s41557-020-0544-y} {\bibfield  {journal} {\bibinfo
  {journal} {Nature Chem.}\ }\textbf {\bibinfo {volume} {12}},\ \bibinfo
  {pages} {891} (\bibinfo {year} {2020})}\BibitemShut {NoStop}%
\bibitem [{\citenamefont {Pfau}\ \emph {et~al.}(2020)\citenamefont {Pfau},
  \citenamefont {Spencer}, \citenamefont {Matthews},\ and\ \citenamefont
  {Foulkes}}]{Pfau:2020}%
  \BibitemOpen
  \bibfield  {author} {\bibinfo {author} {\bibfnamefont {D.}~\bibnamefont
  {Pfau}}, \bibinfo {author} {\bibfnamefont {J.~S.}\ \bibnamefont {Spencer}},
  \bibinfo {author} {\bibfnamefont {A.~G. D.~G.}\ \bibnamefont {Matthews}}, \
  and\ \bibinfo {author} {\bibfnamefont {W.~M.~C.}\ \bibnamefont {Foulkes}},\
  }\href {\doibase 10.1103/PhysRevResearch.2.033429} {\bibfield  {journal}
  {\bibinfo  {journal} {Phys. Rev. Res.}\ }\textbf {\bibinfo {volume} {2}},\
  \bibinfo {pages} {033429} (\bibinfo {year} {2020})}\BibitemShut {NoStop}%
\bibitem [{\citenamefont {Pescia}\ \emph {et~al.}(2024)\citenamefont {Pescia},
  \citenamefont {Nys}, \citenamefont {Kim}, \citenamefont {Lovato},\ and\
  \citenamefont {Carleo}}]{Pescia:2023mcc}%
  \BibitemOpen
  \bibfield  {author} {\bibinfo {author} {\bibfnamefont {G.}~\bibnamefont
  {Pescia}}, \bibinfo {author} {\bibfnamefont {J.}~\bibnamefont {Nys}},
  \bibinfo {author} {\bibfnamefont {J.}~\bibnamefont {Kim}}, \bibinfo {author}
  {\bibfnamefont {A.}~\bibnamefont {Lovato}}, \ and\ \bibinfo {author}
  {\bibfnamefont {G.}~\bibnamefont {Carleo}},\ }\href {\doibase
  10.1103/PhysRevB.110.035108} {\bibfield  {journal} {\bibinfo  {journal}
  {Phys. Rev. B}\ }\textbf {\bibinfo {volume} {110}},\ \bibinfo {pages}
  {035108} (\bibinfo {year} {2024})},\ \Eprint
  {http://arxiv.org/abs/2305.07240} {arXiv:2305.07240 [quant-ph]} \BibitemShut
  {NoStop}%
\bibitem [{\citenamefont {Massella}\ \emph {et~al.}(2020)\citenamefont
  {Massella}, \citenamefont {Barranco}, \citenamefont {Lonardoni},
  \citenamefont {Lovato}, \citenamefont {Pederiva},\ and\ \citenamefont
  {Vigezzi}}]{Massella:2018xdj}%
  \BibitemOpen
  \bibfield  {author} {\bibinfo {author} {\bibfnamefont {P.}~\bibnamefont
  {Massella}}, \bibinfo {author} {\bibfnamefont {F.}~\bibnamefont {Barranco}},
  \bibinfo {author} {\bibfnamefont {D.}~\bibnamefont {Lonardoni}}, \bibinfo
  {author} {\bibfnamefont {A.}~\bibnamefont {Lovato}}, \bibinfo {author}
  {\bibfnamefont {F.}~\bibnamefont {Pederiva}}, \ and\ \bibinfo {author}
  {\bibfnamefont {E.}~\bibnamefont {Vigezzi}},\ }\href {\doibase
  10.1088/1361-6471/ab588c} {\bibfield  {journal} {\bibinfo  {journal} {J.
  Phys. G}\ }\textbf {\bibinfo {volume} {47}},\ \bibinfo {pages} {035105}
  (\bibinfo {year} {2020})},\ \Eprint {http://arxiv.org/abs/1808.00518}
  {arXiv:1808.00518 [nucl-th]} \BibitemShut {NoStop}%
\bibitem [{\citenamefont {Gattobigio}\ \emph {et~al.}(2019)\citenamefont
  {Gattobigio}, \citenamefont {Kievsky},\ and\ \citenamefont
  {Viviani}}]{Gattobigio:2019omi}%
  \BibitemOpen
  \bibfield  {author} {\bibinfo {author} {\bibfnamefont {M.}~\bibnamefont
  {Gattobigio}}, \bibinfo {author} {\bibfnamefont {A.}~\bibnamefont {Kievsky}},
  \ and\ \bibinfo {author} {\bibfnamefont {M.}~\bibnamefont {Viviani}},\ }\href
  {\doibase 10.1103/PhysRevC.100.034004} {\bibfield  {journal} {\bibinfo
  {journal} {Phys. Rev. C}\ }\textbf {\bibinfo {volume} {100}},\ \bibinfo
  {pages} {034004} (\bibinfo {year} {2019})},\ \Eprint
  {http://arxiv.org/abs/1903.08900} {arXiv:1903.08900 [nucl-th]} \BibitemShut
  {NoStop}%
\bibitem [{\citenamefont {K{\"o}nig}(2020)}]{Konig:2019xxk}%
  \BibitemOpen
  \bibfield  {author} {\bibinfo {author} {\bibfnamefont {S.}~\bibnamefont
  {K{\"o}nig}},\ }\href {\doibase 10.1140/epja/s10050-020-00098-9} {\bibfield
  {journal} {\bibinfo  {journal} {Eur. Phys. J. A}\ }\textbf {\bibinfo {volume}
  {56}},\ \bibinfo {pages} {113} (\bibinfo {year} {2020})},\ \Eprint
  {http://arxiv.org/abs/1910.12627} {arXiv:1910.12627 [nucl-th]} \BibitemShut
  {NoStop}%
\bibitem [{\citenamefont {Hildenbrand}\ \emph {et~al.}(2024)\citenamefont
  {Hildenbrand}, \citenamefont {Elhatisari}, \citenamefont {Ren},\ and\
  \citenamefont {Mei{\ss}ner}}]{Hildenbrand:2024ypw}%
  \BibitemOpen
  \bibfield  {author} {\bibinfo {author} {\bibfnamefont {F.}~\bibnamefont
  {Hildenbrand}}, \bibinfo {author} {\bibfnamefont {S.}~\bibnamefont
  {Elhatisari}}, \bibinfo {author} {\bibfnamefont {Z.}~\bibnamefont {Ren}}, \
  and\ \bibinfo {author} {\bibfnamefont {U.-G.}\ \bibnamefont {Mei{\ss}ner}},\
  }\href {\doibase 10.1140/epja/s10050-024-01427-y} {\bibfield  {journal}
  {\bibinfo  {journal} {Eur. Phys. J. A}\ }\textbf {\bibinfo {volume} {60}},\
  \bibinfo {pages} {215} (\bibinfo {year} {2024})},\ \Eprint
  {http://arxiv.org/abs/2406.17638} {arXiv:2406.17638 [nucl-th]} \BibitemShut
  {NoStop}%
\bibitem [{\citenamefont {Hiyama}\ \emph {et~al.}(1996)\citenamefont {Hiyama},
  \citenamefont {Kamimura}, \citenamefont {Motoba}, \citenamefont {Yamada},\
  and\ \citenamefont {Yamamoto}}]{Hiyama:1996gv}%
  \BibitemOpen
  \bibfield  {author} {\bibinfo {author} {\bibfnamefont {E.}~\bibnamefont
  {Hiyama}}, \bibinfo {author} {\bibfnamefont {M.}~\bibnamefont {Kamimura}},
  \bibinfo {author} {\bibfnamefont {T.}~\bibnamefont {Motoba}}, \bibinfo
  {author} {\bibfnamefont {T.}~\bibnamefont {Yamada}}, \ and\ \bibinfo {author}
  {\bibfnamefont {Y.}~\bibnamefont {Yamamoto}},\ }\href {\doibase
  10.1103/PhysRevC.53.2075} {\bibfield  {journal} {\bibinfo  {journal} {Phys.
  Rev. C}\ }\textbf {\bibinfo {volume} {53}},\ \bibinfo {pages} {2075}
  (\bibinfo {year} {1996})}\BibitemShut {NoStop}%
\bibitem [{\citenamefont {Hiyama}\ \emph {et~al.}(1999)\citenamefont {Hiyama},
  \citenamefont {Kamimura}, \citenamefont {Miyazaki},\ and\ \citenamefont
  {Motoba}}]{Hiyama:1999me}%
  \BibitemOpen
  \bibfield  {author} {\bibinfo {author} {\bibfnamefont {E.}~\bibnamefont
  {Hiyama}}, \bibinfo {author} {\bibfnamefont {M.}~\bibnamefont {Kamimura}},
  \bibinfo {author} {\bibfnamefont {K.}~\bibnamefont {Miyazaki}}, \ and\
  \bibinfo {author} {\bibfnamefont {T.}~\bibnamefont {Motoba}},\ }\href
  {\doibase 10.1103/PhysRevC.59.2351} {\bibfield  {journal} {\bibinfo
  {journal} {Phys. Rev. C}\ }\textbf {\bibinfo {volume} {59}},\ \bibinfo
  {pages} {2351} (\bibinfo {year} {1999})}\BibitemShut {NoStop}%
\bibitem [{\citenamefont {Yao}\ \emph {et~al.}(2011)\citenamefont {Yao},
  \citenamefont {Li}, \citenamefont {Hagino}, \citenamefont {Win},
  \citenamefont {Zhang},\ and\ \citenamefont {Meng}}]{Yao:2011wp}%
  \BibitemOpen
  \bibfield  {author} {\bibinfo {author} {\bibfnamefont {J.~M.}\ \bibnamefont
  {Yao}}, \bibinfo {author} {\bibfnamefont {Z.~P.}\ \bibnamefont {Li}},
  \bibinfo {author} {\bibfnamefont {K.}~\bibnamefont {Hagino}}, \bibinfo
  {author} {\bibfnamefont {M.~T.}\ \bibnamefont {Win}}, \bibinfo {author}
  {\bibfnamefont {Y.}~\bibnamefont {Zhang}}, \ and\ \bibinfo {author}
  {\bibfnamefont {J.}~\bibnamefont {Meng}},\ }\href {\doibase
  10.1016/j.nuclphysa.2011.08.006} {\bibfield  {journal} {\bibinfo  {journal}
  {Nucl. Phys. A}\ }\textbf {\bibinfo {volume} {868-869}},\ \bibinfo {pages}
  {12} (\bibinfo {year} {2011})},\ \Eprint {http://arxiv.org/abs/1104.3200}
  {arXiv:1104.3200 [nucl-th]} \BibitemShut {NoStop}%
\bibitem [{\citenamefont {Lonardoni}\ and\ \citenamefont
  {Pederiva}(2017)}]{Lonardoni:2017uuu}%
  \BibitemOpen
  \bibfield  {author} {\bibinfo {author} {\bibfnamefont {D.}~\bibnamefont
  {Lonardoni}}\ and\ \bibinfo {author} {\bibfnamefont {F.}~\bibnamefont
  {Pederiva}},\ }\href@noop {} {\  (\bibinfo {year} {2017})},\ \Eprint
  {http://arxiv.org/abs/1711.07521} {arXiv:1711.07521 [nucl-th]} \BibitemShut
  {NoStop}%
\bibitem [{\citenamefont {Piarulli}\ \emph {et~al.}(2018)\citenamefont
  {Piarulli} \emph {et~al.}}]{Piarulli:2017dwd}%
  \BibitemOpen
  \bibfield  {author} {\bibinfo {author} {\bibfnamefont {M.}~\bibnamefont
  {Piarulli}} \emph {et~al.},\ }\href {\doibase 10.1103/PhysRevLett.120.052503}
  {\bibfield  {journal} {\bibinfo  {journal} {Phys. Rev. Lett.}\ }\textbf
  {\bibinfo {volume} {120}},\ \bibinfo {pages} {052503} (\bibinfo {year}
  {2018})},\ \Eprint {http://arxiv.org/abs/1707.02883} {arXiv:1707.02883
  [nucl-th]} \BibitemShut {NoStop}%
\end{thebibliography}%
\end{document}